\documentclass[epj]{svjour}
\usepackage{graphicx}
\begin{document}

\title{Extended $s$-wave pairing symmetry on the triangular lattice heavy fermion system}
\author{Lan Zhang\inst{1}\thanks{\emph{Present address:} zhanglan14@lzu.edu.cn} \and Yu-Feng Wang\inst{1} \and Yin Zhong\inst{1} \and Hong-Gang Luo \inst{1},\inst{2}
\thanks{\emph{Present address:} luohg@lzu.edu.cn}%
}                     
\institute{Center for Interdisciplinary Studies $\&$ Key Laboratory for
Magnetism and Magnetic Materials of the MoE, Lanzhou University, Lanzhou 730000, China \and Beijing Computational Science Research Center, Beijing 100084, China}
\date{Received: date / Revised version: date}
%
\abstract{
We investigate the pairing symmetry of the Kondo-Heisenberg model on triangular lattice, which is believed to capture the core competition of Kondo screening and local magnetic exchange interaction in heavy electron compounds. On the dominant background of the heavy fermion state, the introduction of the Heisenberg antiferromagnetic interaction ($J_H$) leads to superconducting pairing instability. Depending on the strength of the interactions, it is found that the pairing symmetry prefers an extended $s$-wave for small $J_H$ and high conduction electron density but a chiral $d_{x^2-y^2}+id_{xy}$-wave for large $J_H$ and low conduction electron density, which provides a phase diagram of pairing symmetry from the calculations of the ground-state energy. The transition between these two pairing symmetries is found to be first-order. Furthermore, we also analyze the phase diagram from the pairing strengths and find that the phase diagram obtained is qualitatively consistent with that based on the ground-state energy. In addition, we propose an effective single-band BCS Hamiltonian, which is able to describe the low-energy thermodynamic behaviors of the heavy fermion superconducting states. These results further deepen the understanding of the antiferromagnetic interaction which results in a geometric frustration for the model studied. Our work may provide a possible scenario to understand the pairing symmetry of the heavy fermion superconductivity, which is one of active issues in very recent years.
} 
\maketitle
\section{Introduction}
The Kondo lattice model is at the heart position to understand the ground-state property of the heavy fermion compounds determined by the competition between the Kondo effect and the Ruderman-Kittel-Kasuya-Yosida (RKKY) interaction (i.e., Doniach picture) both obtained in principle from the intrasite Kondo coupling.\cite{Stewart2001,Hilbertv2007,Doni77PBC-,Lacroix1979} However, among the Ce-heavy fermion compounds, some exhibit antiferromagnetic long-range order,\cite{Regnault1988,Pierre1990,Chattopadhyay1994,Flouquet1995} which motivated Glesias, Lacroix, and Coqblin to revisit the ``Doniach picture" by including explicitly the intersite Heisenberg interaction.\cite{Arispe1995255,Iglesias1996160,Lacroix1997503} The revisited version of Doniach picture is able to capture the magnetic nature of those Ce-compounds.\cite{Igle97PRB-RDd}

The introduction of the Heisenberg coupling leads to many novel phenomena.\cite{Thomas2014,Montiel2014,Asadzadeh2014,Liu2012,LIU2014,Otsuki2015} Among those phenomena, the occurrence of the superconductivity is apparently beyond the physics of the original Kondo lattice.\cite{Coleman1989,Xavier2003} In a recent paper,\cite{LIU2014} Liu, Zhang, and Yu found that the heavy fermion state of the Kondo-Heisenberg model on the square lattice is superconducting unstable and it favors a d-wave pairing symmetry, which is reminiscent of the d-wave symmetry of the Cu-based high-temperature superconductors resulted from the low-energy effective t-J model on the square CuO$_2$ plane.\cite{Anderson1987} Due to the puzzling pairing mechanism of the cuprates, the search of unconventional superconductors becomes one of central topics in condensed matter physics in recent three decades.\cite{Pickett1989,Harlingen1995,Basov2005,Lee2006,Norman2011}

On the other hand, the water-intercalated sodium cobaltates Na$_{x}$Co$_{2}\cdot y$H$_{2}$O was found in layered metal oxides Na$_{x}$Co$_{2}$ through a chemical oxidation process and the families of organic charge-transfer salts $\kappa$-(ET)$_2$X and\\ Pd(dmit)$_2$ are two examples to exhibit unconventional superconductivity.\cite{Williams1991,Takada2003,Kiesel2013,Sasaki2002,Rau2011,Kanoda1997} The essential physics of these materials lies on the geometrically frustrated triangular lattice.\cite{Kino1996,Kino1998,Shimizu2003,Itou2007,Tocchio2009} The interplay of the electronic correlations and the geometric frustration is the physical origin of many exotic emergent phenomena like quantum spin liquid states.\cite{Yamashita2010,Balents2010} In fact, it is also believed that the interplay could lead to a chiral pairing state which breaks parity and time-reversal symmetry.\cite{Kiesel2012,Nandkishore2012} These recent developments motivate us to further explore the superconductivity of the Kondo-Heisenberg model (KHM) on the triangular lattice.

The triangular lattice has a $C_{6v}$ rotational symmetry, and it allows a doubly degenerate $E_2$ representation of the superconducting order parameters with $d_{x^2-y^2}$ and $d_{xy}$ degenerate states.\cite{Kiesel2013} This indicates that the KHM on the triangular lattice has a chiral superconducting ground-state and should exhibit a $d_{x^2-y^2} + id_{xy} (d+id)$ pairing symmetry.

In the present work, we employ the large-N mean-field approach to study this model. As expected, we find that the superconducting state with chiral $d+id$ pairing symmetry can be possible for large Heisenberg antiferromagnetic interaction ($J_H$) and low conduction electron density ($n_c$) from the calculations of the ground-state energy. Surprisingly, for small $J_H$ and high $n_c$, the pairing symmetry is found to favor an extended $s$-wave. Thus we obtain a phase diagram of pairing symmetry of the model. Furthermore, a qualitatively similar phase diagram has also been obtained from the calculations of the pairing strengths, which signals that the local Heisenberg antiferromagnetic interaction drives pairing instability with Fermi surface topology. This provides a physically intuitive understanding of the existence of the extended $s$-wave pairing symmetry in such a heavy fermion system. It is also found that the phase transition between the extended $s$-wave and the chiral $d+id$-wave is a first-order when tuning $J_H$ and $n_c$. In addition, we analyze the characteristics of the spectra and the density of states of quasiparticles for different pairing symmetries. In the superconducting state, an effective single-band BCS Hamiltonian is proposed to describe the superconducting properties such as superfluid density of the model. Finally, one notes that the issue of the pairing symmetry of heavy fermion systems becomes very active due to some experimental and theoretical advances\cite{Kittaka2014,Kim2015,Erten2015,Masuda2015} and our work provides a possible scenario in understanding the pairing symmetry of these heavy fermion systems.

The paper is organized as follows. In Sec.\ref{sec:1}, we introduce the Kondo-Heisenberg lattice model under the large-N mean-field theory. In Sec.\ref{sec:2}, we present the results and give the phase diagram of the pairing symmetries both from the calculations of the ground-state energy and the pairing strengths. The characteristics of the quasiparticle spectra, the density of states, and the pairing functions have also presented. Some discussion on an effective single-band BCS model is also given. Finally, Sec.\ref{sec:3} is devoted to a brief conclusion and perspective.

\section{Model and mean-field approach}\label{sec:1}
The Hamiltonian of the KHM on the triangular lattice can be written as
\begin{eqnarray}\label{eq:A1}
&& H = -t\sum_{\left\langle ij\right\rangle,\sigma}c^{\dag}_{i\sigma}c_{j\sigma}+t_{1}\sum_{\left\langle\left\langle ij\right\rangle\right\rangle,\sigma}c^{\dag}_{i\sigma}c_{j\sigma} -\mu\sum_{i\sigma}c^{\dag}_{i\sigma}c_{i\sigma} \nonumber\\
&& \hspace{1cm} + J_K\sum_i \textbf{S}_{i}\cdot\textbf{s}_{i} + J_{H}\sum_{\langle ij\rangle}\textbf{S}_{i}\cdot\textbf{S}_{j},
\end{eqnarray}
where $c^{\dag}_{i\sigma}(c_{i\sigma})$ denotes the creation (annihilation) operator of the conduction electrons with spin $\sigma$. The first line in Eq. (\ref{eq:A1}) describes the hoppings of the conduction electrons and $\mu$ is the chemical potential. The notations of $\langle \cdot \rangle$ and $\langle\langle \cdot \rangle\rangle$ represent the nearest-neighbor (NN) and the next-nearest-neighbor (NNN) hopping respectively. The introduction of the NNN hopping term is to avoid the occasional nesting. The $J_{K}$ (the coupling strength) term in the second line denotes the Kondo coupling between the localized f-electrons and conduction electrons. The $\textbf{S}_{i}= \frac{1}{2}\sum_{\alpha\beta}f_{i\alpha}\tau_{\alpha\beta}f_{i\beta}$ is the spin operator of localized f-electron with the local constraint $\sum_{\sigma} f^{\dag}_{i\sigma}f_{i\sigma}=1$, the $\textbf{s}_{i}=\frac{1}{2}\sum_{\sigma\sigma^{\prime}}c^{\dag}_{i,\sigma} \tau_{\sigma\sigma^{\prime}} c_{i,\sigma^{\prime}}$ is the spin operator of the conduction electrons, and $\tau$ is the Pauli matrix. The last $J_H$ (the interaction strength) term is the Heisenberg exchange interaction introduced by Coleman and Andrei to explore possible spin-liquid-stabilized unconventional pairing states.\cite{Coleman1989} It has also been used by Glesias \textit{et al.} to consider the antiferromagnetic long-range order in some Ce-based heavy fermion compounds.\cite{Arispe1995255,Iglesias1996160,Lacroix1997503}

The standard large-N mean-field method\cite{Read1983} is sufficient to deal with the KHM on the triangular lattice. Since we consider the assumption that the pairing originates from the antiferromagnetic exchange interaction, the spin-singlet pairing is expected to have lower energy compared to the spin-triplet pairing. In addition, the possible pairing symmetry should be compatible with the lattice symmetry of the triangular lattice. Thus, we consider two kinds of pairing order parameters, namely, the extended $s$-wave pairing $\Delta_{ij}=\Delta_{0}$ and the chiral $d+id$-wave paring $\Delta_{ij}=\Delta_{0}e^{2i\theta_{ij}}$ breaking the parity and time-reversal symmetry, where $\theta_{ij}$ is the angle between the hopping direction and the horizon. The extended $s$-wave has a uniform phase $\theta_{ij}=0$, while the chiral $d+id$-wave has three possible values, namely, $\theta_{ij} = (0, \frac{\pi}{3}, \frac{2\pi}{3})$.

By the large-N mean-field, the Heisenberg antiferromagnetic interaction can be decoupled by the particle-particle and particle-hole channels, i.e., $\Delta_{ij} =-\langle f^{\dag}_{i,\uparrow}f^{\dag}_{j,\downarrow}-f^{\dag}_{i,\downarrow}f^{\dag}_{j,\uparrow}\rangle$ and $\chi_{ij}=-\langle f^{\dag}_{i,\uparrow}f_{j,\uparrow}+f^{\dag}_{i,\downarrow}f_{j,\downarrow}\rangle$, and the Kondo screening channel $V=\langle f^{\dag}_{i,\uparrow}c_{i,\uparrow}+f^{\dag}_{i,\downarrow}c_{i,\downarrow}\rangle$. \cite{Liu2012} As a result, $\textbf{S}_{i}\cdot\textbf{S}_{j} =\frac{1}{2}[\Delta_{ij}(f^{\dag}_{i\uparrow}f^{\dag}_{j\downarrow}-f^{\dag}_{i\downarrow}f^{\dag}_{j\uparrow})
+H.c.]+\frac{|\Delta_{ij}|^2}{2}=\frac{1}{2}[\chi_{ij}(f^{\dag}_{i\uparrow}f_{j\uparrow}
+f^{\dag}_{i\downarrow}f_{j\downarrow})+H.c.]+\frac{|\chi_{ij}|^2}{2}$. Based on these mean-field formulations, Eq.(\ref{eq:A1}) can be rewritten in the $k$-space as follows
\begin{eqnarray}\label{eq:A10}
&&H=\sum_{k}\Psi^{\dag}_{k}\left(
                           \begin{array}{cccc}
                             \varepsilon_{k}-\mu & 0 & -\frac{J_{K}V}{2} & 0 \\
                             0 & -\varepsilon_{k}+\mu & 0 & \frac{J_{K}V}{2} \\
                             -\frac{J_{K}V}{2} & 0 & \chi_{k} & J_{H}\Delta_{k} \\
                             0 & \frac{J_{K}V}{2} & J_{H}\Delta^{\star}_{k} & -\chi_{k}\\
                           \end{array}
                         \right)\Psi_{k}\nonumber\\
&&+\sum_{k}\left(\varepsilon_{k}-\mu\right)
+N_{S}\left(\frac{J_{K}V^{2}}{2}+\frac{3J_{H}\Delta_{0}^{2}}{2}+\frac{3J_{H}\chi^{2}}{2}\right),
\end{eqnarray}
where $\Psi^{\dag}_{\textbf{k}}=\left(c^{\dag}_{\textbf{k}\uparrow}, c_{-\textbf{k}\downarrow}, f^{\dag}_{\textbf{k}\uparrow}, f_{-\textbf{k}\downarrow}\right)$ is a four-component Nambu spinor, and $\Delta_{k}$ is the pairing function corresponding to the extended $s$-wave or the chiral $d+id$-wave pairing in momentum space. $\varepsilon_{k} = -2t\gamma_{k}+2t_{1}\left[\textrm{cos}\left(\sqrt{3}k_y\right)\right.$ $\left.+2\cos\left(\frac{\sqrt{3}}{2}k_y\right)\cos\left(\frac{3}{2}k_x\right)\right]$ with $\gamma_{k} =  2\cos\left(\frac{\sqrt{3}}{2}k_y\right)\cos\left(\frac{k_x}{2}\right)$ $+\cos\left(k_x\right) $ is the single-particle energy of the non-interacting conduction electrons. $\chi_{k}=J_{H}\chi\gamma_{k}+\lambda$ is the kinetic energy of the $f$-electrons, denoting the dispersion of real f-electron band. The parameter $\chi = |\chi_{ij}|$ and $\lambda$ is the Lagrange multiplier due to the local constraint of the localized electrons. $N_S$ is the number of sites in the lattice. The pairing order parameters for the extended $s$-wave and the chiral $d+id$-wave read,
\begin{eqnarray}\label{eq:A11}
&&\Delta^{s}_{k}=\Delta_{0}\left[\cos(k_{x})
+2\cos\left(\frac{\sqrt{3}k_{y}}{2}\right)\cos\left(\frac{k_{x}}{2}\right)\right],\\
&&\Delta^{d+id}_{k}=\Delta_{0}\left[\cos(k_{x})
-\cos\left(\frac{k_{x}}{2}\right)\cos\left(\frac{\sqrt{3}k_{y}}{2}\right)\right.\nonumber\\
&&\hspace{2cm}\left.-i\sqrt{3}\sin\left(\frac{k_{x}}{2}\right)\sin\left(\frac{\sqrt{3}k_{y}}{2}\right)\right].
\end{eqnarray}
Diagonalizing the mean-field Hamiltonian, the quasiparticle energy spectra have two independent branches, which read
\begin{equation}\label{eq:A12}
E^{\pm}_{k}=\sqrt{E_{k1}\pm\sqrt{E^{2}_{k1}-E^{2}_{k2}}},
\end{equation}
where $E_{k1}=\frac{1}{2}\left[\xi_{k}^{2}+J^{2}_{H}|{\Delta_{k}}|^{2}+\chi_{k}^{2}+\frac{1}{2}J^{2}_{K}V^{2}\right]$, and $E_{k2}=\sqrt{J^{2}_{H}|{\Delta_{k}}|^{2}\xi_{k}^{2}+X_k^{2}}$.

At zero temperature, the ground-state energy reads
\begin{equation}\label{eq:A13}
E_{g}=\frac{1}{N_{S}}\sum_{k}\left(\varepsilon_{k}-\mu-E^{0}_{k}\right)+\frac{J_{K}V^{2}}{2}
+\frac{3J_{H}\Delta^{2}_{0}}{2}+\frac{3J_{H}\chi^{2}}{2},
\end{equation}
where $E^{0}_{k}=E^{+}_{k}+E^{-}_{k}=\sqrt{2\left(E_{k1}+E_{k2}\right)}$, $\xi_{k}=\varepsilon_{k}-\mu$, and $X_k = \chi_{k}\xi_{k}-\frac{J^{2}_{K}V^{2}}{4}$. Minimizing the ground-state energy, one obtains a set of the self-consistent equations to determine the parameters $\chi, \Delta_{0}, V$, $\lambda$, and the chemical potential $\mu$:
\begin{eqnarray}\label{eq:A14}
&&\frac{1}{N_{S}}
\sum_{k}\frac{1}{E_{k}}\left[1+\frac{\xi^{2}_{k}}{E_{k2}}\right]|{\Delta_{k}}|^{2}=\frac{3\Delta^{2}_{0}}{J_{H}},\label{mf-1}\\
&&\frac{1}{N_{S}}
\sum_{k}\frac{1}{E_{k}}\left[1-\frac{X_k}{E_{k2}}\right]=\frac{2}{J_{K}},\\
&&\frac{1}{N_{S}}
\sum_{k}\frac{1}{E_{k}}\left[\chi_{k}+\frac{\xi_{k}X_k}{E_{k2}}\right]=0,\\
&&\frac{1}{N_{S}}\sum_{k}
\frac{1}{E_{k}}\left[\chi_{k}+\frac{\xi_{k}X_k}{E_{k2}}\right]\gamma_{k}
=3\chi,\\
&&\frac{1}{N_{S}}\sum_{k}\frac{1}{E_{k}}\left[\xi_{k}
+\frac{\chi_{k}X_k+\xi_{k}J^{2}_{H}\Delta^{2}_{k}}{E_{k2}}\right]=1-n_{c}, \label{mf-5}
\end{eqnarray}
where the last equation is due to the constraint of the concentration of conduction electrons.

\section{Unconventional superconductivity}\label{sec:2}
\subsection{pairing symmetry}
\begin{figure}[tdp]
\centering
\includegraphics[width = 0.485\columnwidth]{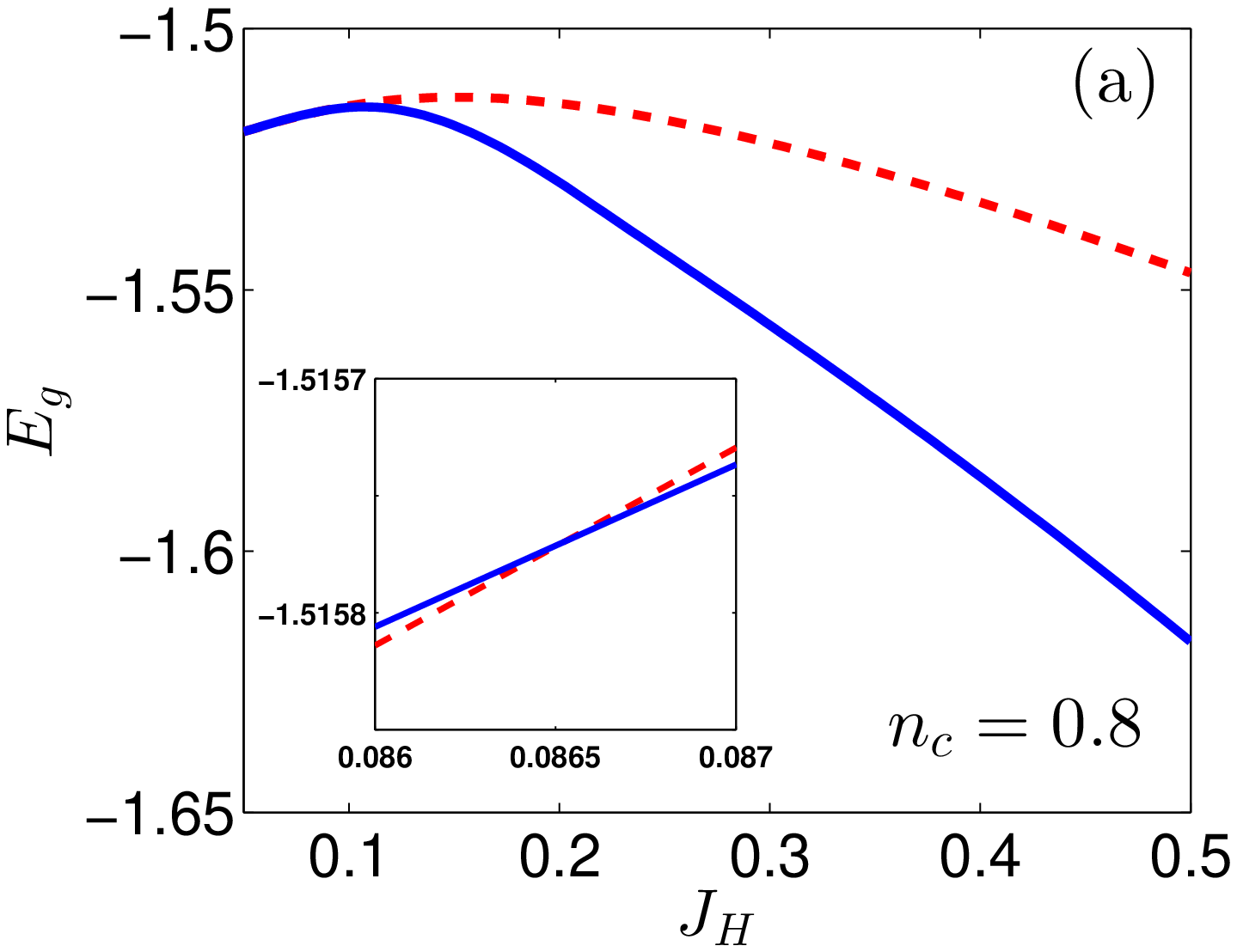}
\includegraphics[width = 0.485\columnwidth]{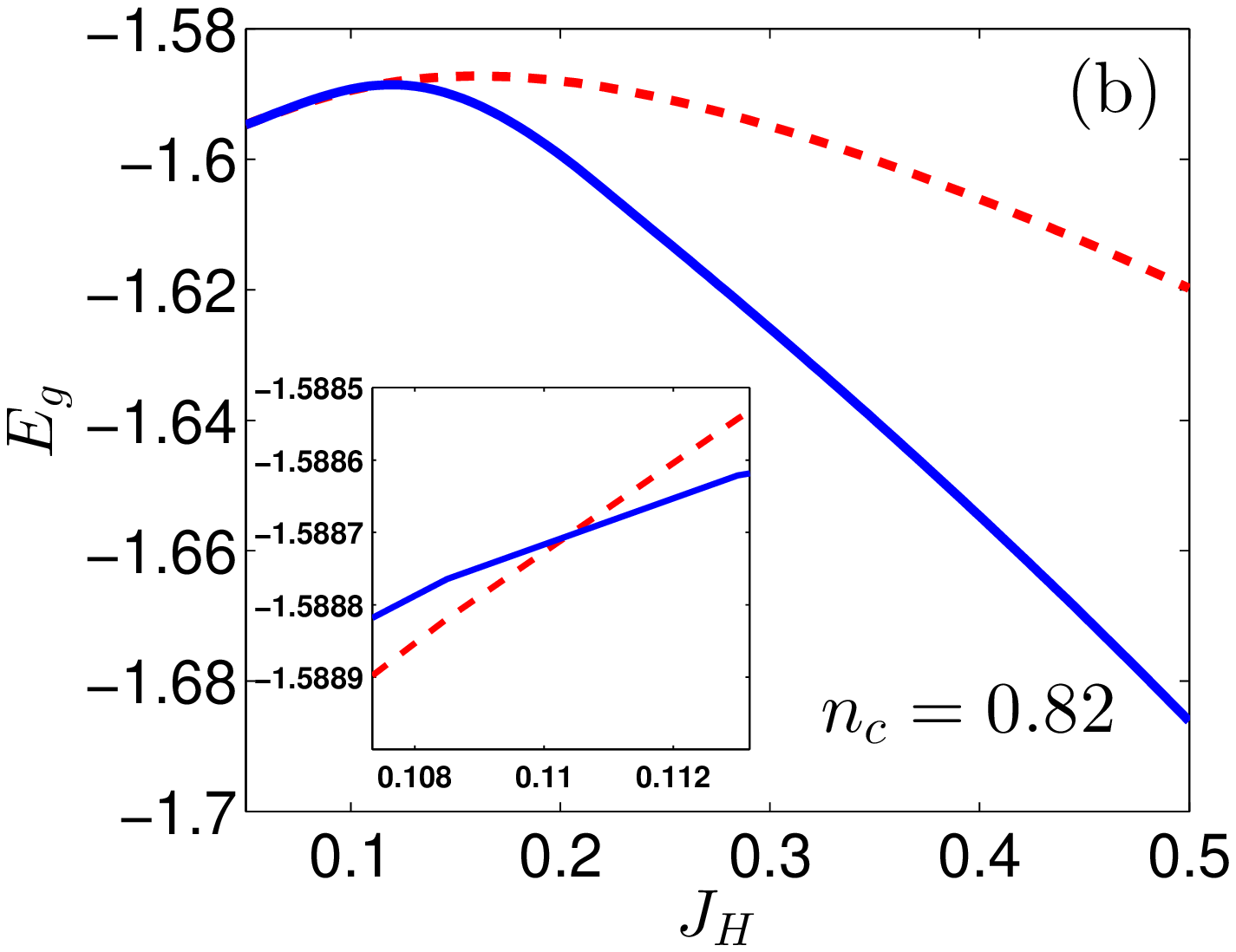}
\includegraphics[width = 0.485\columnwidth]{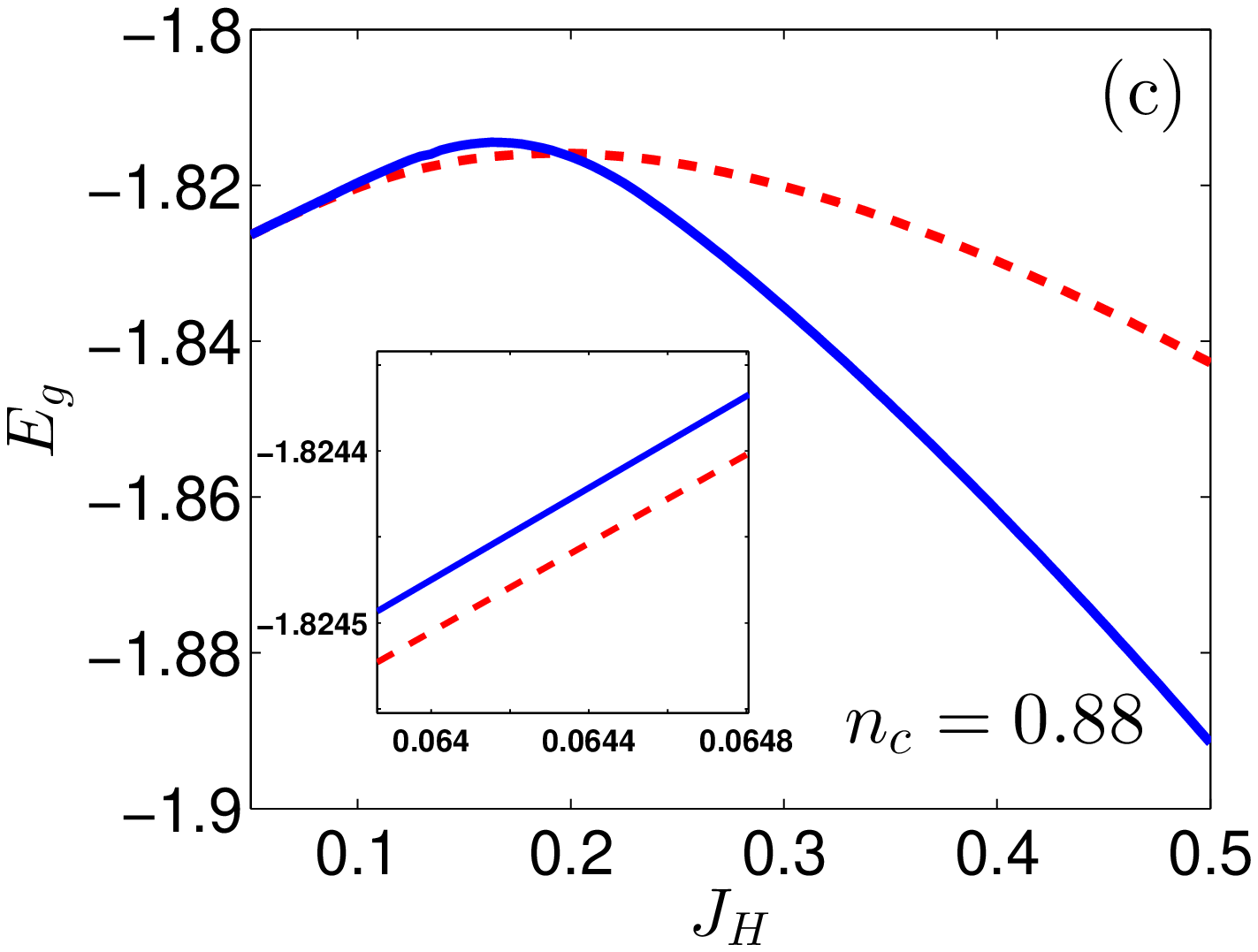}
\includegraphics[width = 0.485\columnwidth]{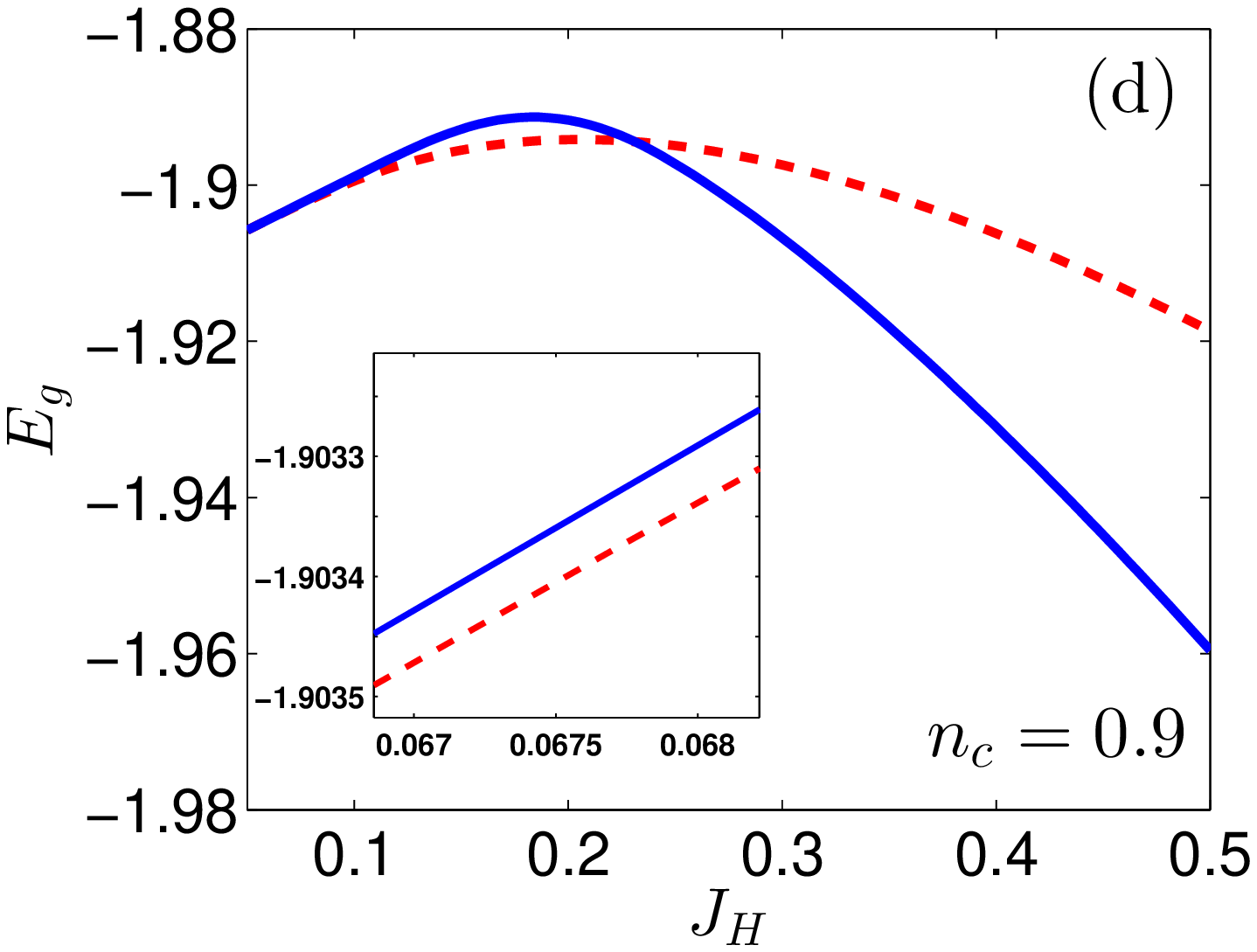}
\caption{\label{fig:pic4} The ground-state energy $E_{g}$ as a function of the Heisenberg antiferromagnetic interaction $J_H$ in unit of $t$ (here and hereafter the unit of energy is taken as $t$) for different conduction electron concentrations. The red dashed lines correspond to the extended $s$-wave pairing and the blue solid ones correspond to the chiral $d+id$-wave pairing. The insets show the cases of small $J_H$. The parameters used are $t_{1}/t=0.3$, $J_{K}/t=2.5$.}
\end{figure}

It is straightforward to solve self-consistently the mean-field equations given by Eqs. (\ref{mf-1})-(\ref{mf-5}) by assuming the corresponding pairing symmetry, namely, the chiral $d+id$ (blue solid line) and the extended $s$-wave (red dashed line). First of all, we discuss the ground-state energies for these two different pairing order parameters. For different conduction electron density $n_c$, the ground-state energy is plotted as a function of the Heisenberg antiferromagnetic interaction $J_H$, as shown in Fig. \ref{fig:pic4}. For all $n_c$, it is found that irrespective of the pairing symmetry, the ground-state energy shows non-monotonic behavior of $J_H$, which indicates the competition between the Heisenberg antiferromagnetic interaction and the Kondo screening. Comparing with different pairing symmetry, it is found that for large $J_H$ the system with the chiral $d+id$-wave pairing has a lower energy than that with the extended $s$-wave pairing. Surprisingly, with decreasing of $J_H$, the two lines go smoothly across, and as a result, the system with the extended $s$-wave pairing has a lower energy than that with the chiral $d+id$-wave pairing. For very small $J_H$, the insets in Fig. \ref{fig:pic4} show that the ground-state energy with the extended $s$-wave pairing is still lower than that with the chiral $d+id$-wave pairing, although their energies are very close. This result indicates that for small $J_H$ the system favors the extended $s$-wave pairing, rather than the commonly believed $d$-wave pairing in such heavy fermion systems.

It is well known that the superconducting pairing symmetry plays an important role in understanding of the pairing mechanism. For example, for the weak-coupling superconductivity, e.g., the conventional BCS superconductors, the pairing symmetry is an $s$-wave, in which electron pairings due to the glue of phonon have a large weight on the on-site pairing amplitude in real space. For the strong-coupling superconductivity, e.g., high T$_c$ cuprates, the situation is quite different. The strong Coulomb repulsion disfavors the on-site pairing but it drives the antiferromagnetic fluctuations. Thus, in high T$_c$ cuprates the anisotropic pairing shows the $d$-wave symmetry. For the heavy fermion systems, many experimental and theoretical results, e.g., Ref. \cite{Ishida1999}, leaded to a commonly expectation that the heavy fermion superconductors should have a $d$-wave pairing symmetry since the superconductivity is very close proximity to the antiferromagnetic region. However, some very recent works indicated that the $s$-wave scenario is possible in the heavy fermion systems, which will be left for a brief review in the final section. Here we continue to discuss the characteristic behaviors of the heavy fermion superconducting states at different pairing symmetry.

We first check the quasi-particle energy spectra obtained. One notes that nodal points will appear if $E_{k2}=0$, which means that $\left|\Delta_{k}\right|^{2}=0$ and $\xi_{k}=\frac{J^{2}_{K}V^{2}}{4\chi_{k}}$ need to satisfy at the same time. However, it is obvious that these two conditions are difficult to satisfy at the same time irrespective of the pairing symmetry assumptions of the extended $s$- or the chiral $d+id$-type pairing.

\begin{figure}[tdp]
\centering
\includegraphics[width = 0.485\columnwidth]{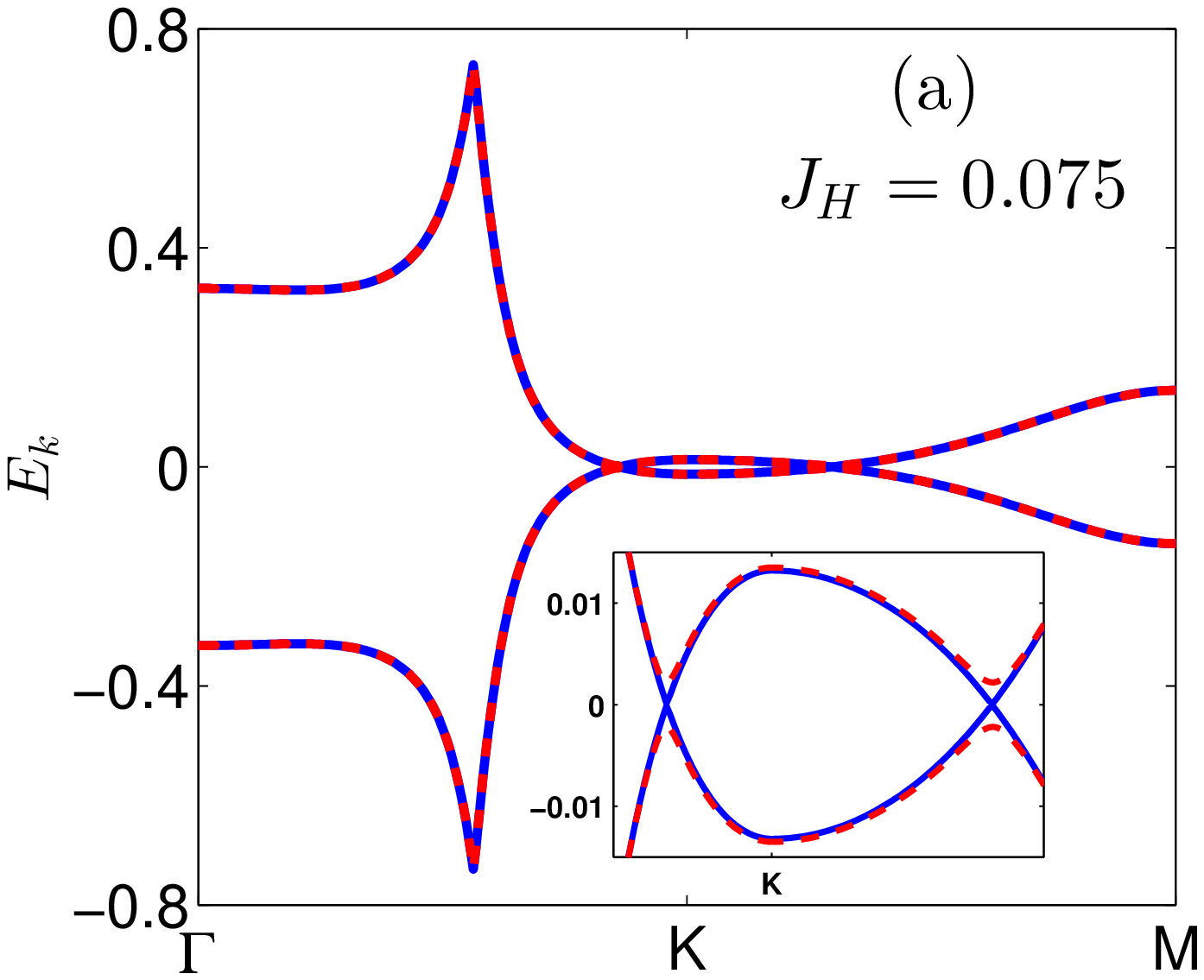}
\includegraphics[width = 0.485\columnwidth]{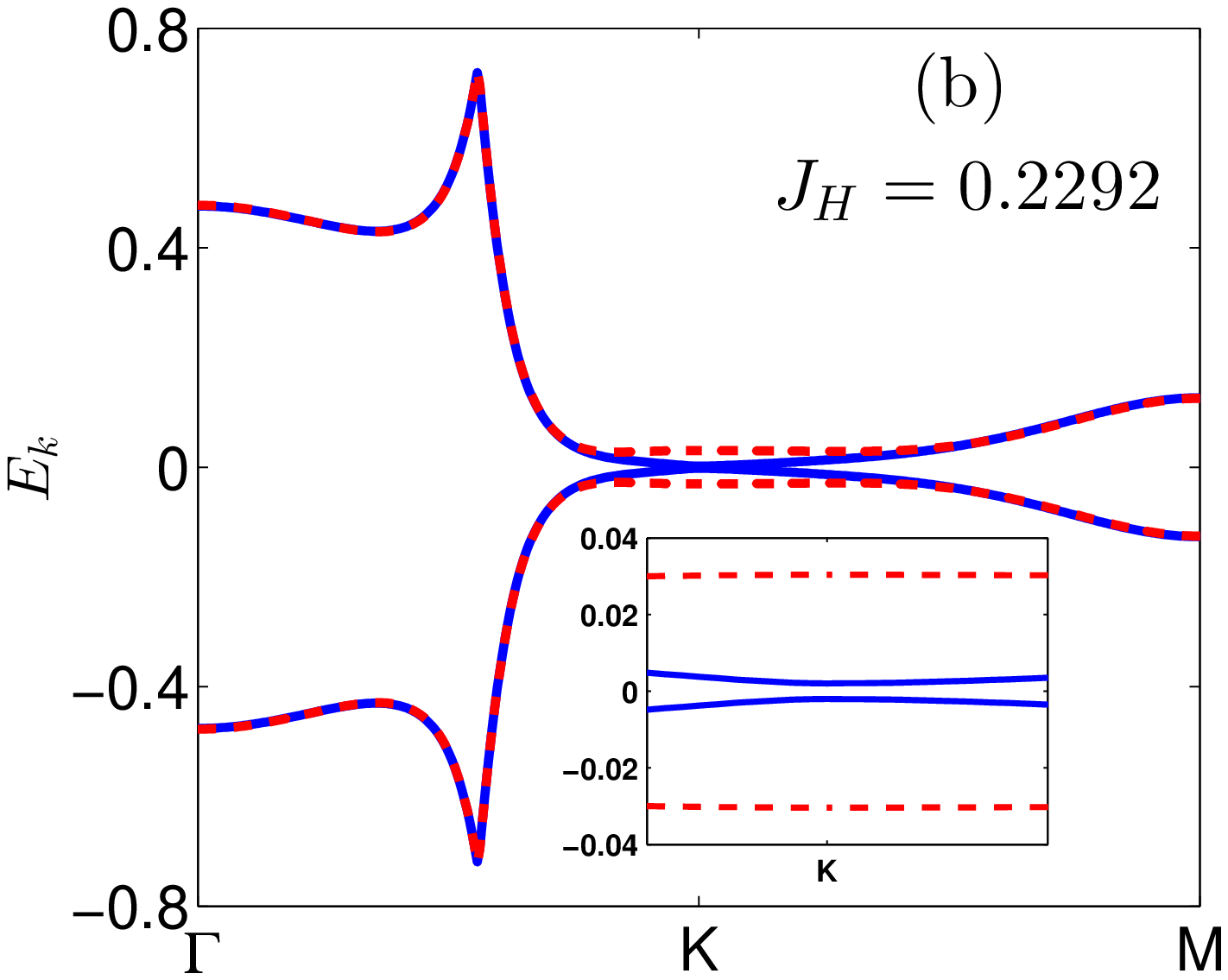}
\includegraphics[width = 0.485\columnwidth]{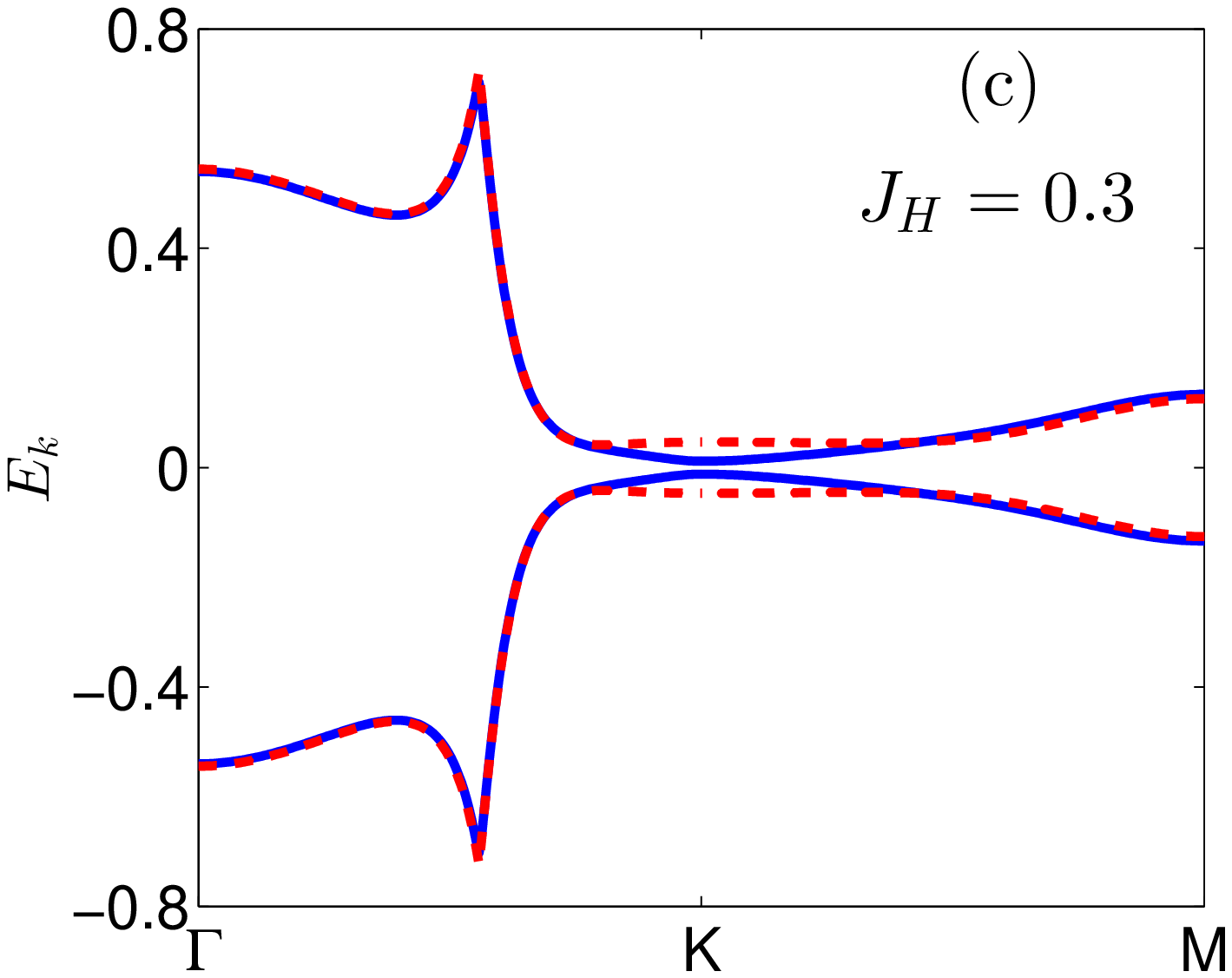}
\includegraphics[width = 0.485\columnwidth]{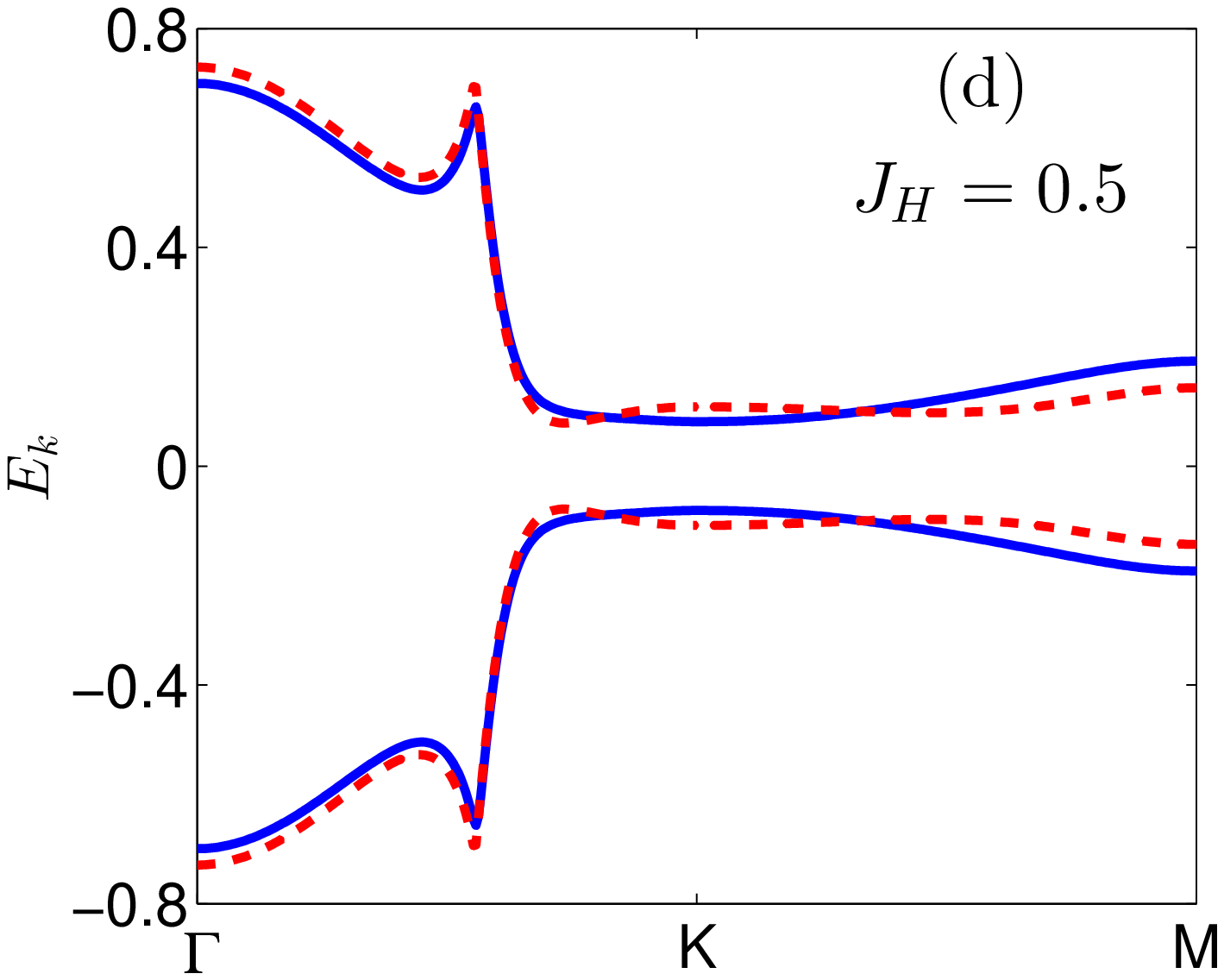}
\caption{\label{fig:pic5} The quasi-particle energy band structure for the branch of $E_k^-$ near the Fermi surface. The other branch of $E_k^+$ (not shown here) is away from the Fermi surface. The red dashed lines correspond to the extended $s$-wave pairing and the blue solid ones correspond to the chiral $d+id$-wave pairing. Here $n_c = 0.9$ and the other parameters used are the same as those in Fig. \ref{fig:pic4}.}
\end{figure}

It is found that the higher energy band branch of $E_k^+$ is away from the Fermi surface, which shows trivial feature with different $J_H$. However, the energy band $E_k^-$ is close to the Fermi surface. The numerical calculation of the quasi-particle energy spectrum $E^{-}_{k}$ are shown in Fig. \ref{fig:pic5} for the $d+id$ (blue solid line) and $s$-wave(red dashed line) pairing symmetry. They shows interesting features with different $J_H$. For small $J_H = 0.075$, the extended $s$-wave energy spectrum of $E^-_k$ has two points very close to the Fermi surface which shows two negligible energy gap. In contrast, the chiral $d+id$-wave is almost gapless, as presented in the inset of Fig. \ref{fig:pic5}(a). With increasing $J_H$ up to $J_H = 0.2292$, the gap of the extended $s$-wave pairing increases, and a sizable gap begins to open for the chiral $d+id$-wave pairing, as shown in Fig. \ref{fig:pic5}(b). This is the cross point of the ground-state energy for the extended $s$-wave and the chiral $d+id$-wave pairing symmetry at $n_c = 0.9$, as shown in Fig. \ref{fig:pic4}(d). Further increasing $J_H$, both energy gaps increase dramatically, and in this regime, the system always favors the chiral $d+id$-wave pairing symmetry when comparing the ground-state energies of two kinds of pairing symmetry. From these discussion, one can conclude that the system favors an open but smaller energy gap in heavy fermion superconducting state. A similar energy spectra feature has also been observed in the KHM on the square lattice.\cite{Liu2012} Thus, it is reasonable that the extended $s$-wave pairing should also exist in the case of the square lattice.\cite{LIU2014}

\begin{figure}[tdp]
\centering
\includegraphics[width = 0.485\columnwidth]{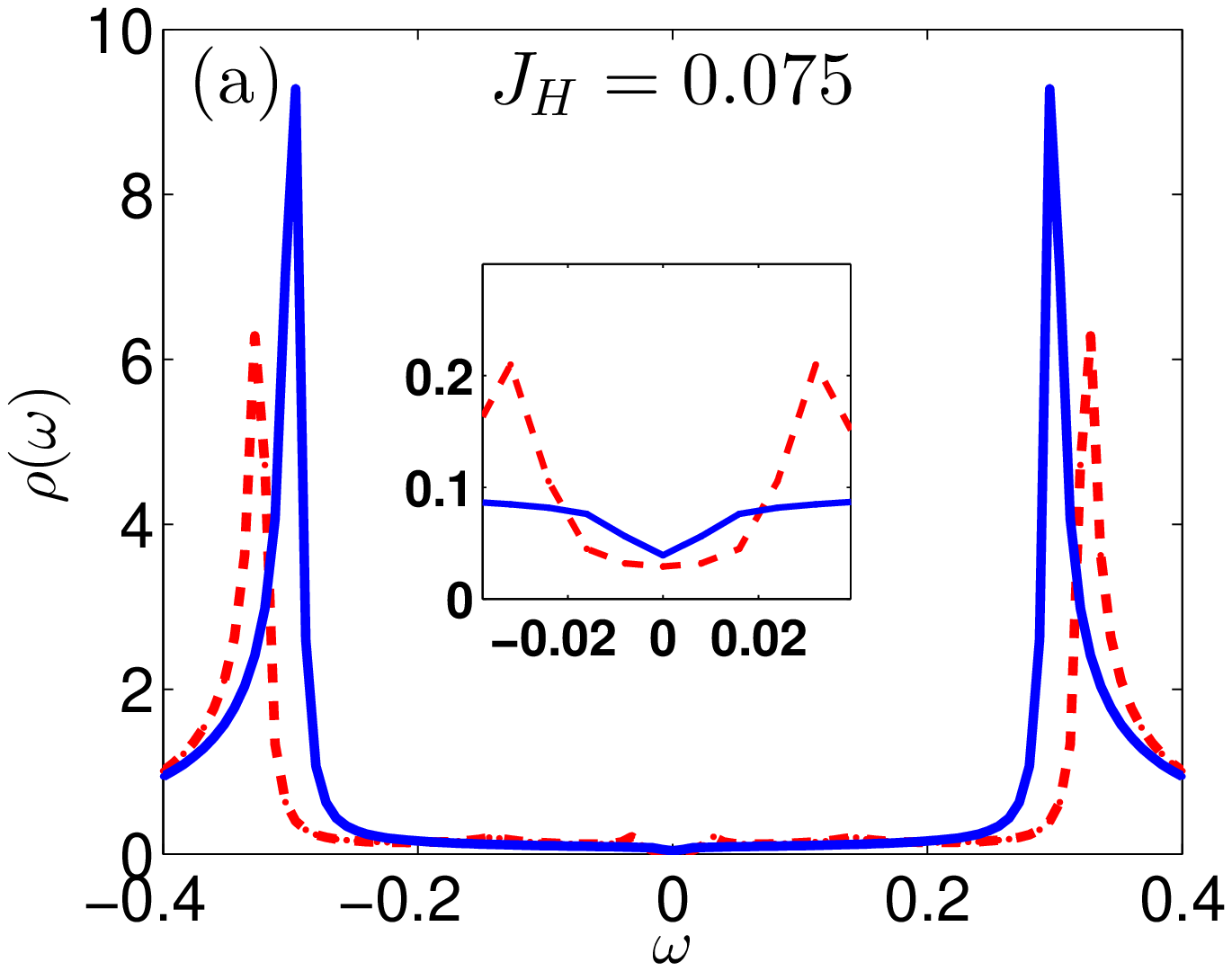}
\includegraphics[width = 0.485\columnwidth]{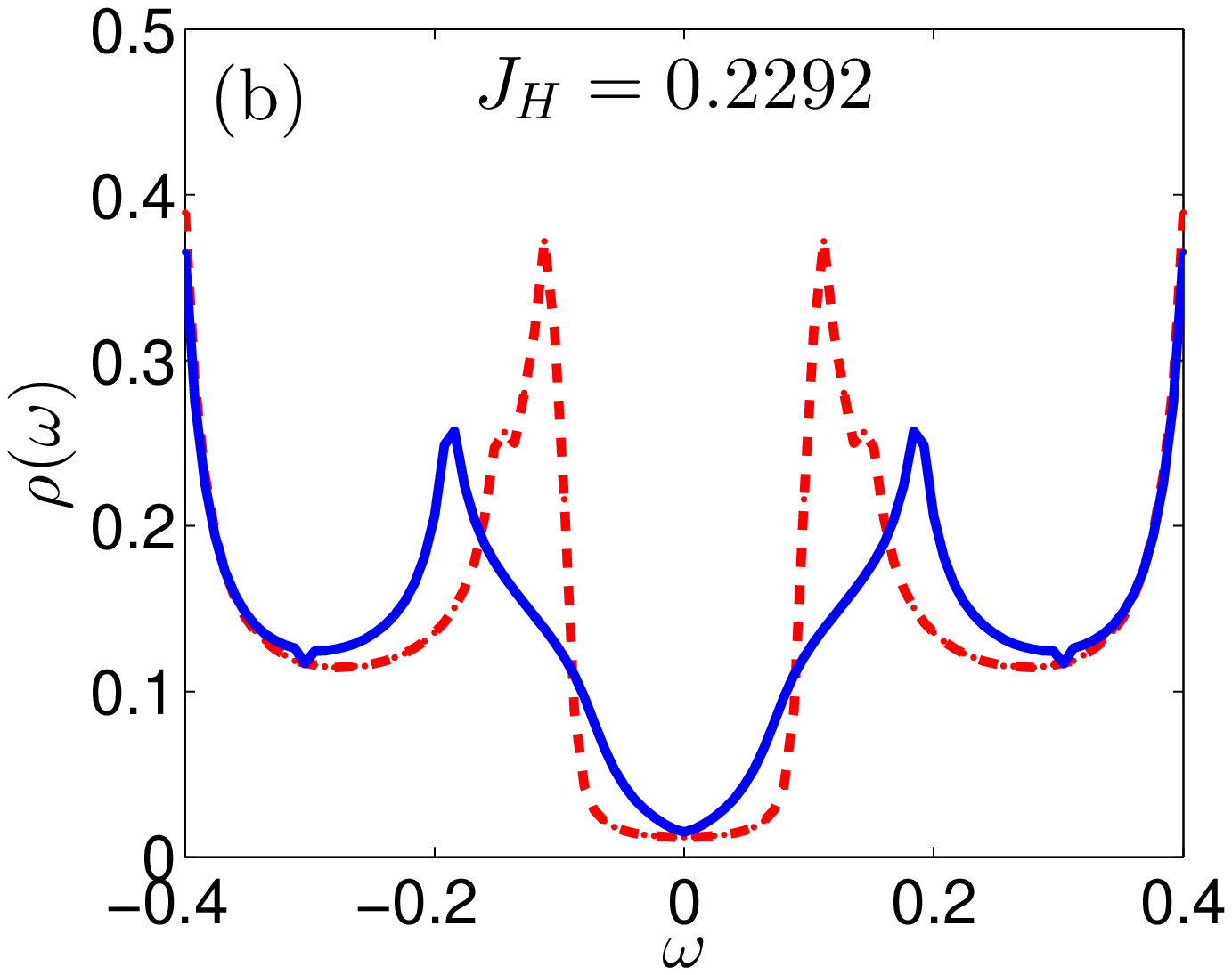}
\includegraphics[width = 0.485\columnwidth]{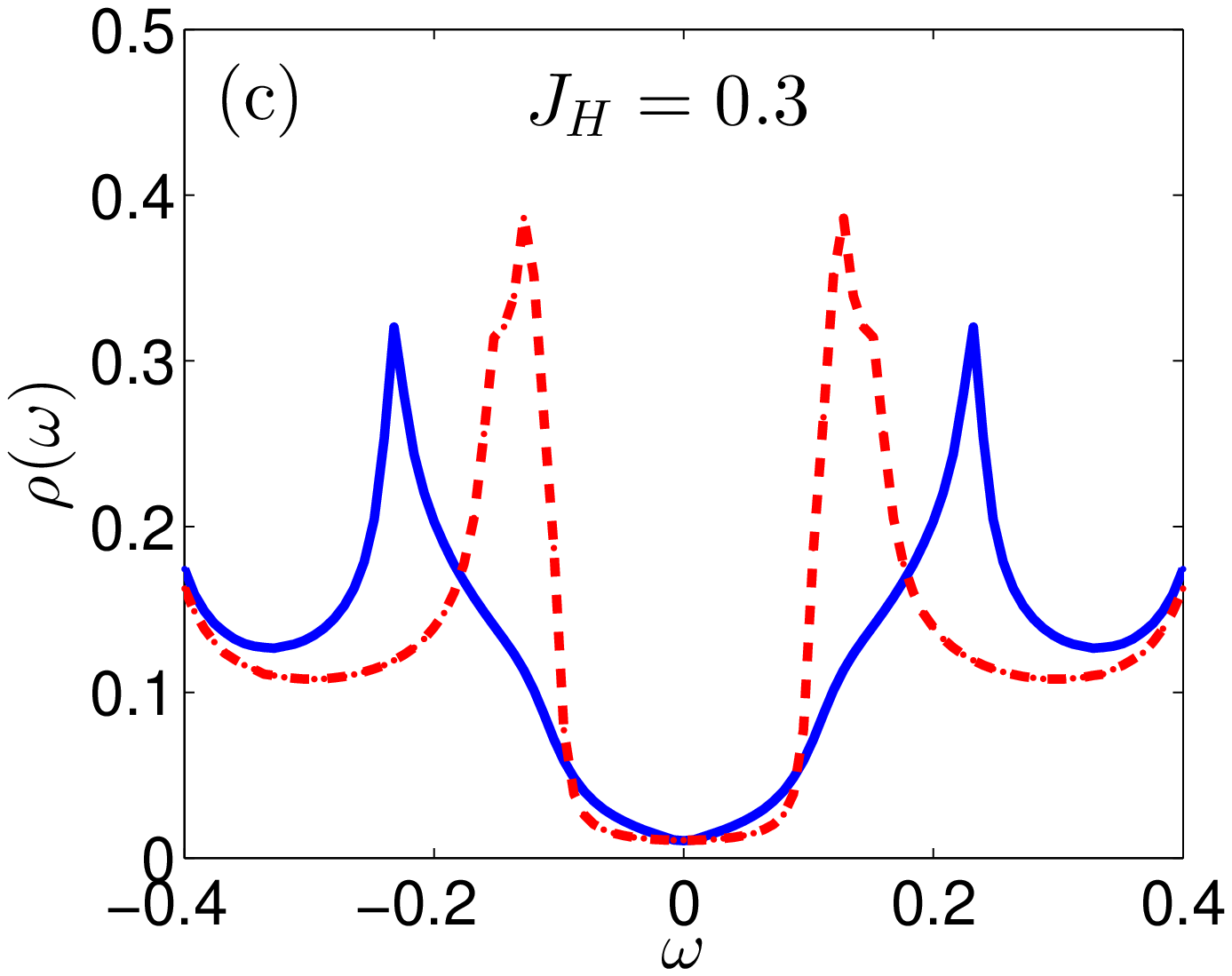}
\includegraphics[width = 0.485\columnwidth]{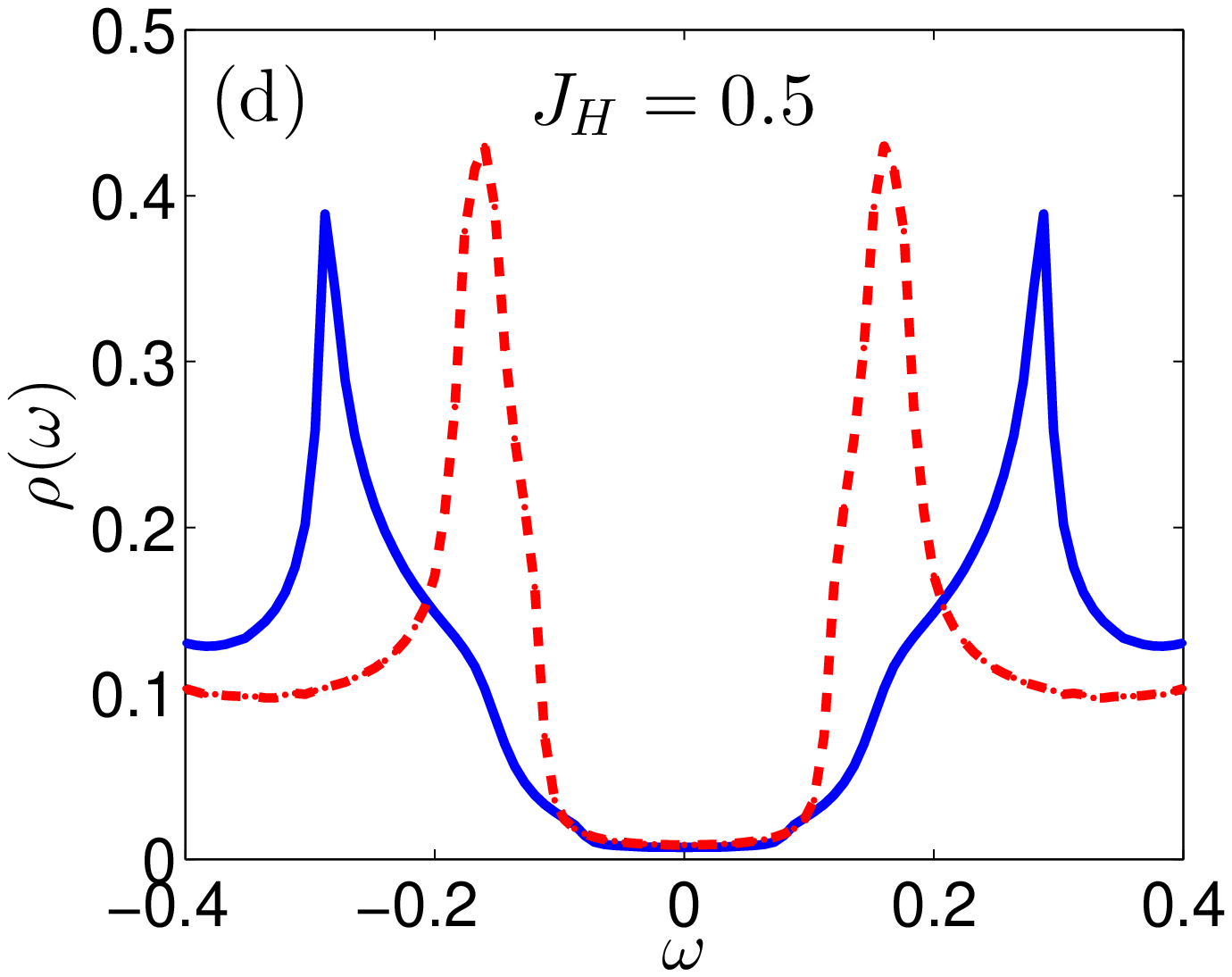}
\caption{\label{fig:pic7} The conduction electrons density of state for different $J_H$. The parameters used are the same as those in Fig. \ref{fig:pic5}. }
\end{figure}

The transition of the pairing symmetry can also be reflected by the conduction electron density of states calculated by using Green's function $G(k,\tau)=-\langle T_{\tau}c_{k\sigma}(\tau)c^{\dag}_{k\sigma}(0)\rangle$, as shown in Fig. \ref{fig:pic7}. For small $J_H$, only the extended $s$-wave pairing shows a sizable energy gap, but for the chiral $d+id$-wave pairing assumption the energy gap is absent (Note: a finite density of states around the Fermi level for both cases is due to finite broadening in the calculations). This result is related to the above observation that the superconducting states in the KHM always favors a finite energy gap, thus have the extended $s$-wave pairing for small $J_H$. With increasing $J_H$ up to 0.2292 for $n_c = 0.9$[see, Fig. \ref{fig:pic7}(b)], the energy gap for the extended $s$-wave pairing assumption becomes more and more larger, and the chiral $d+id$-wave begins to open a sizable gap. After the quantum phase transition, for large $J_H$, the system begins to favor the chiral $d+id$-wave with a finite energy gap. A finite energy gap for the chiral $d+id$-wave pairing is consistent with the nature of the chiral $d+id$-wave pairing.\cite{Kiesel2013} In addition, one also notes that in the chiral $d+id$-wave pairing case, the superconducting coherent peaks are obviously absent, which is sharp contrast to the case of the extended $s$-wave pairing.

\begin{figure}[!t]
\centering
\includegraphics[width = 0.32\columnwidth]{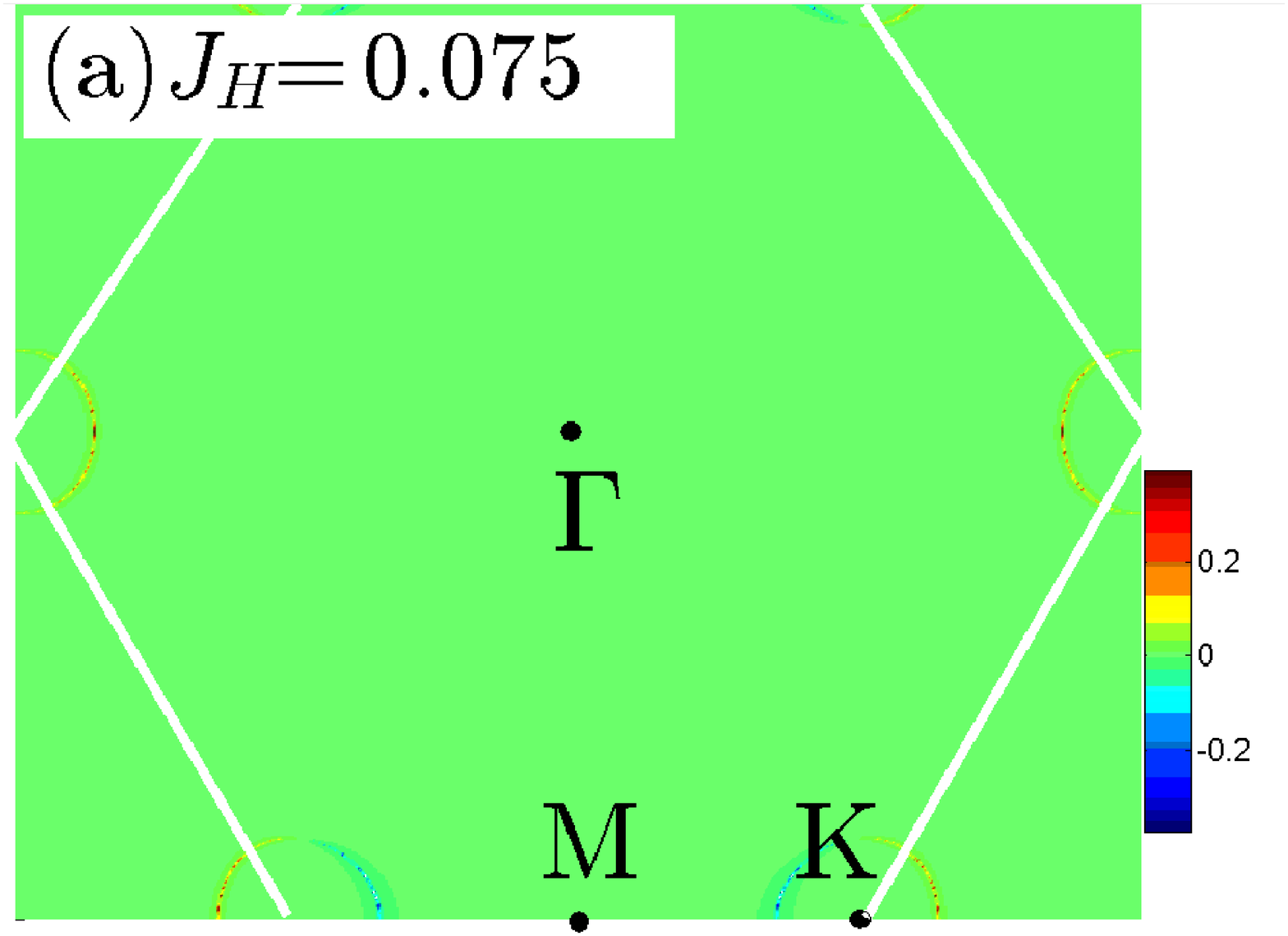}
\includegraphics[width = 0.32\columnwidth]{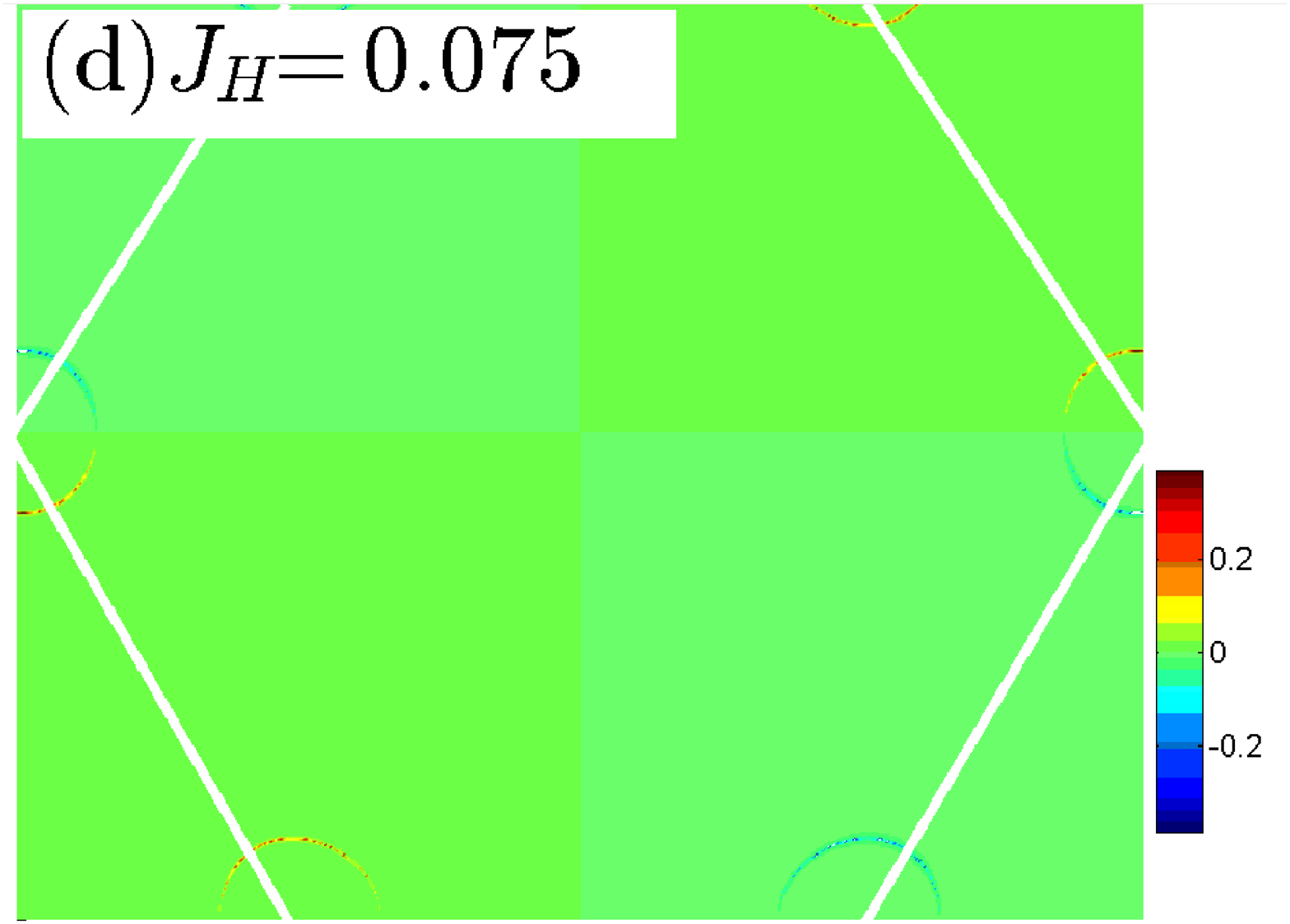}
\includegraphics[width = 0.32\columnwidth]{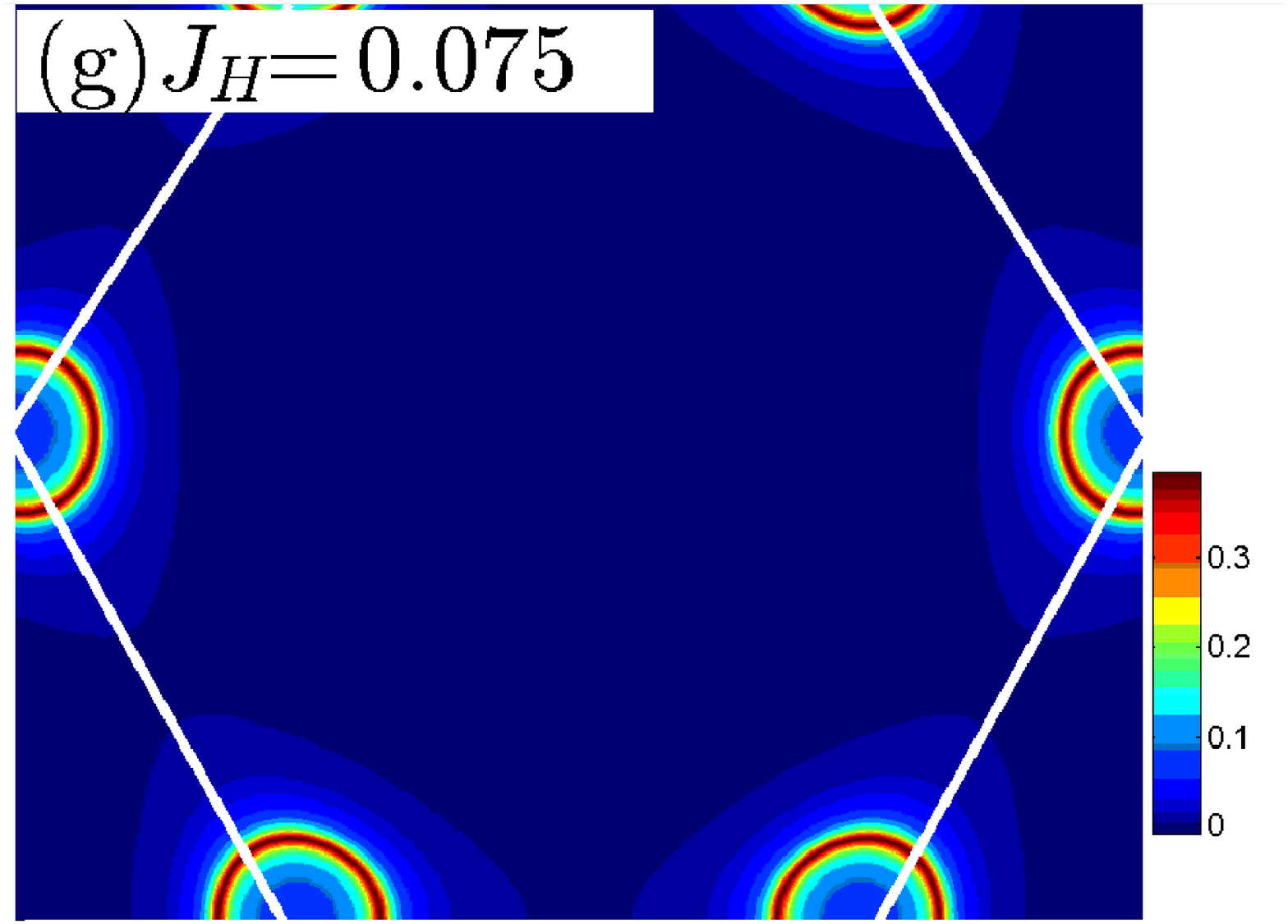}
\includegraphics[width = 0.32\columnwidth]{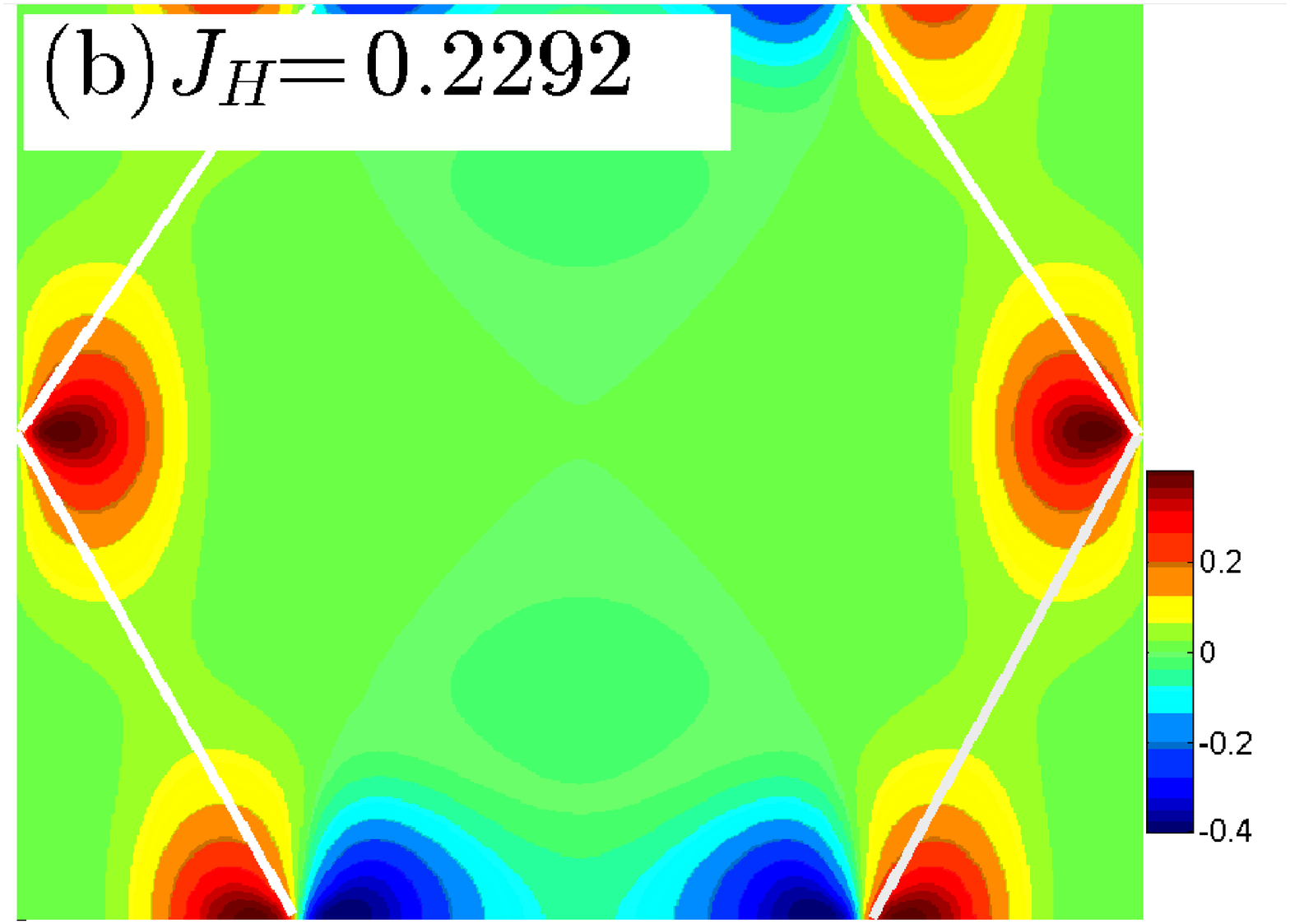}
\includegraphics[width = 0.32\columnwidth]{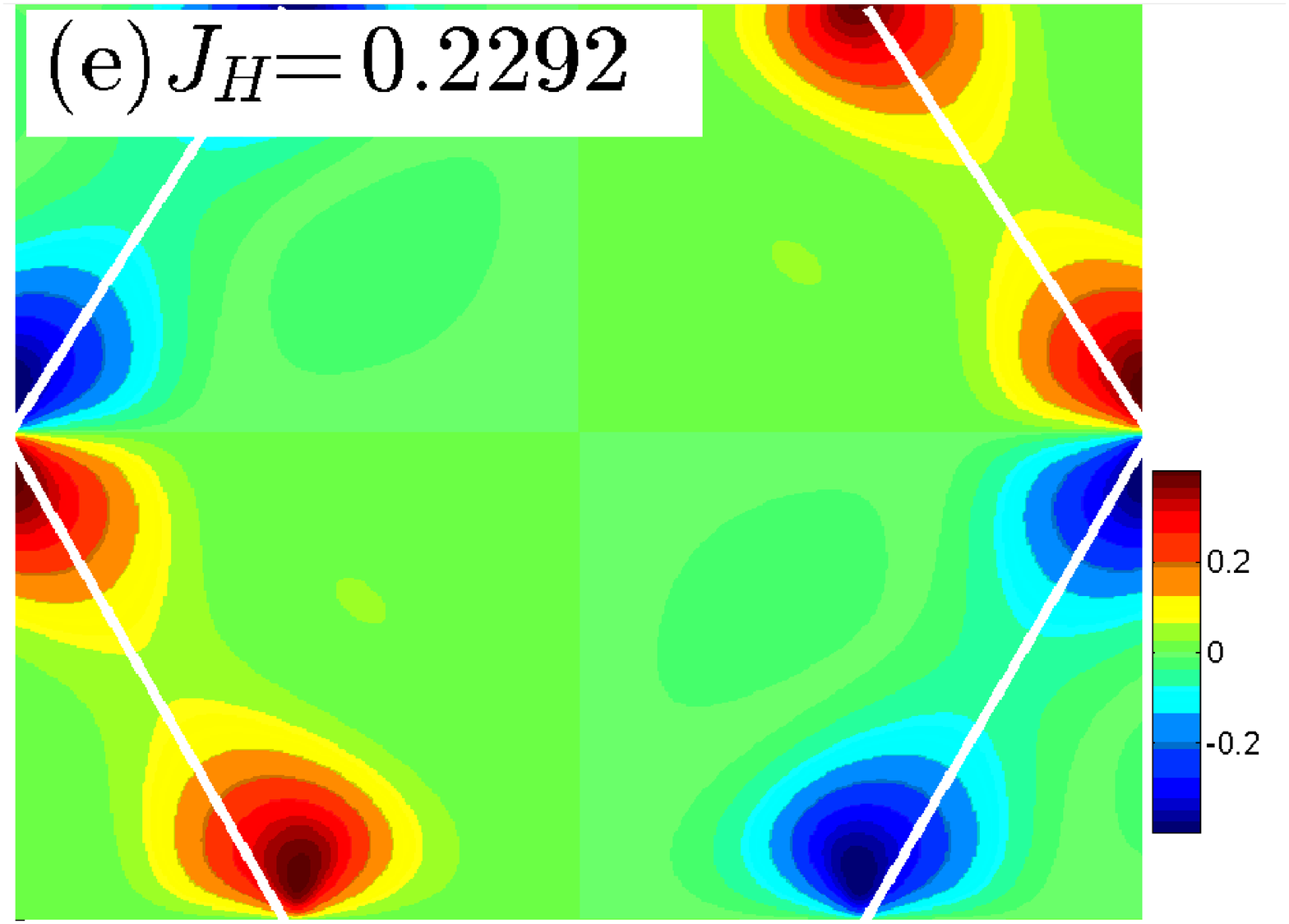}
\includegraphics[width = 0.32\columnwidth]{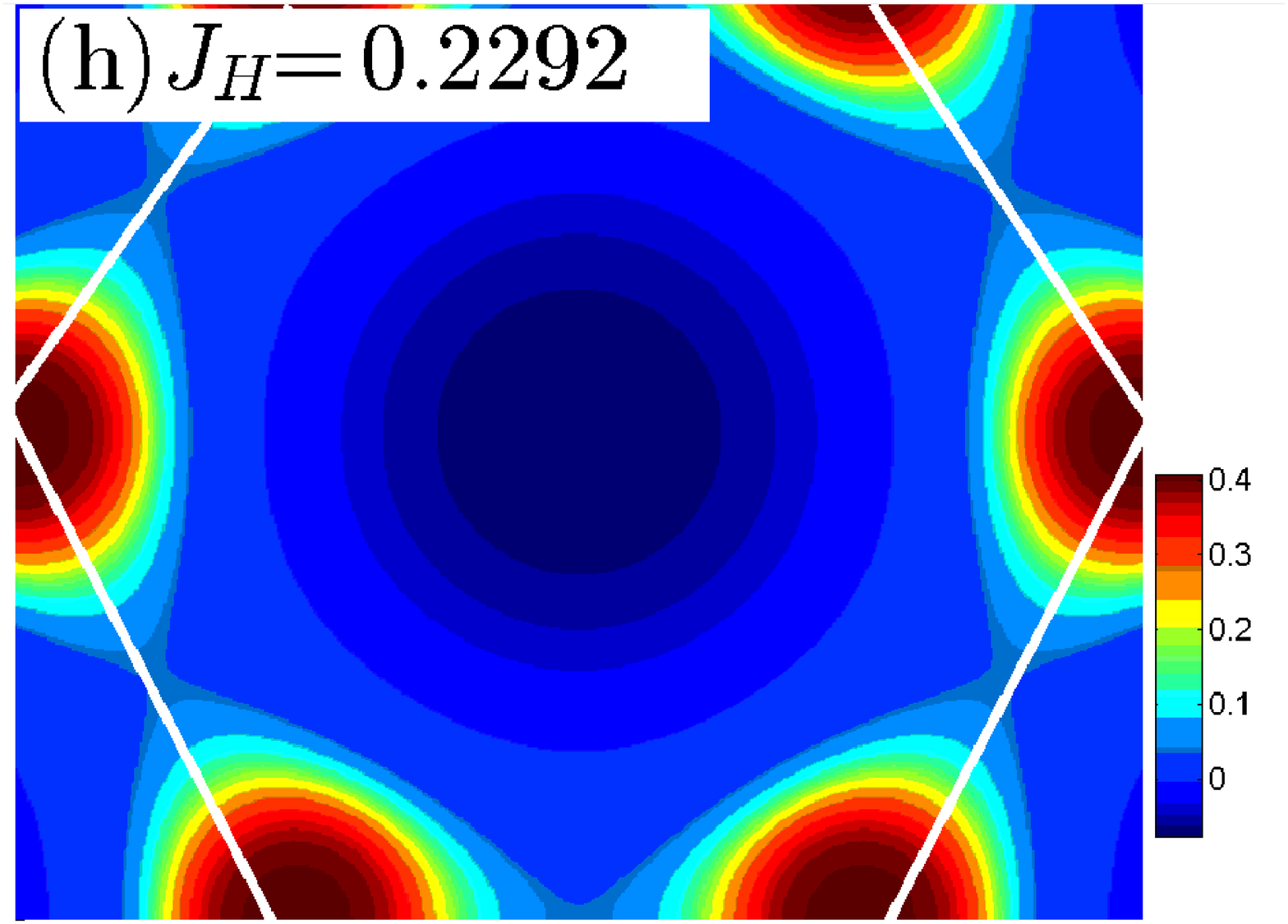}
\includegraphics[width = 0.32\columnwidth]{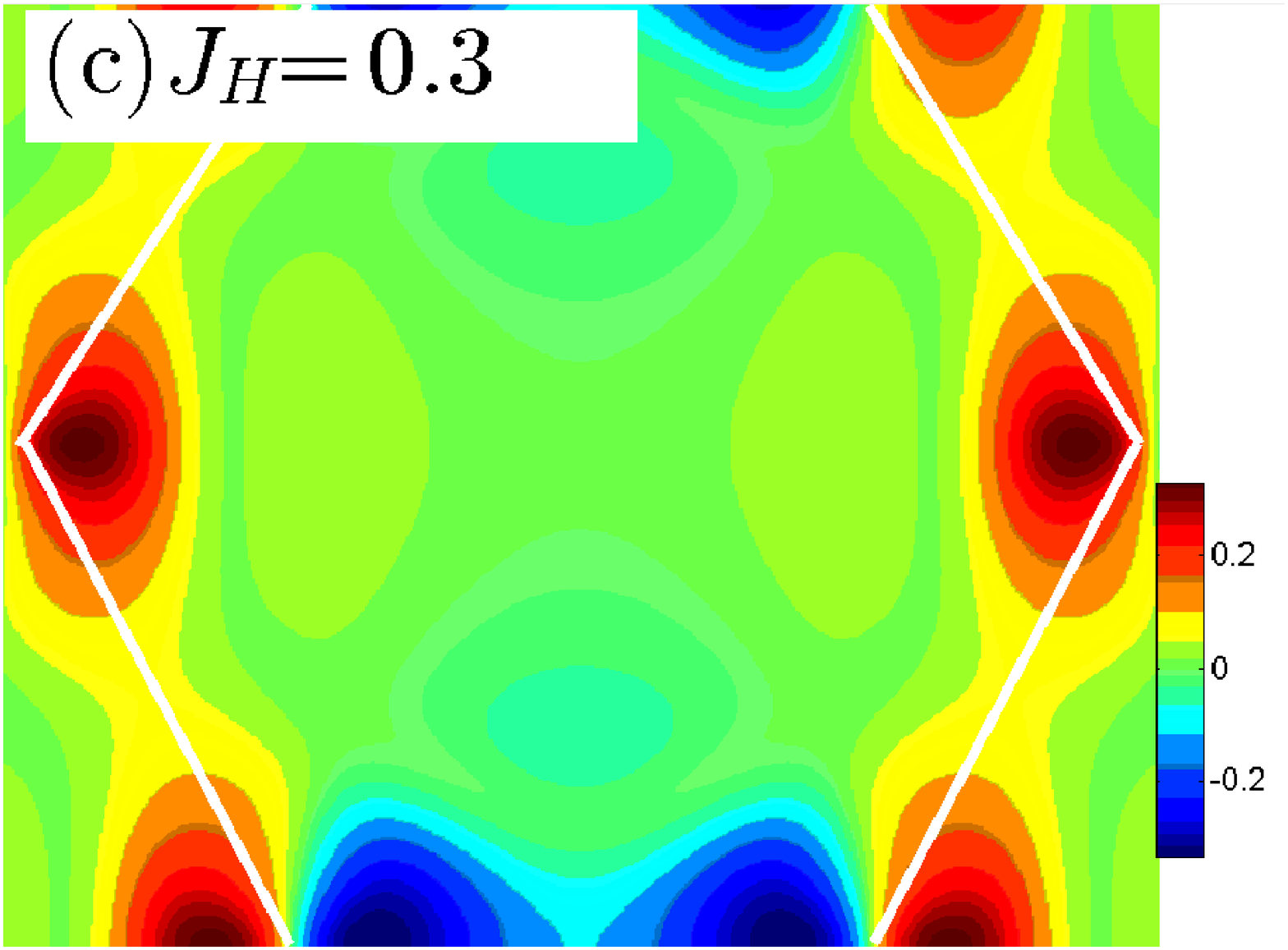}
\includegraphics[width = 0.32\columnwidth]{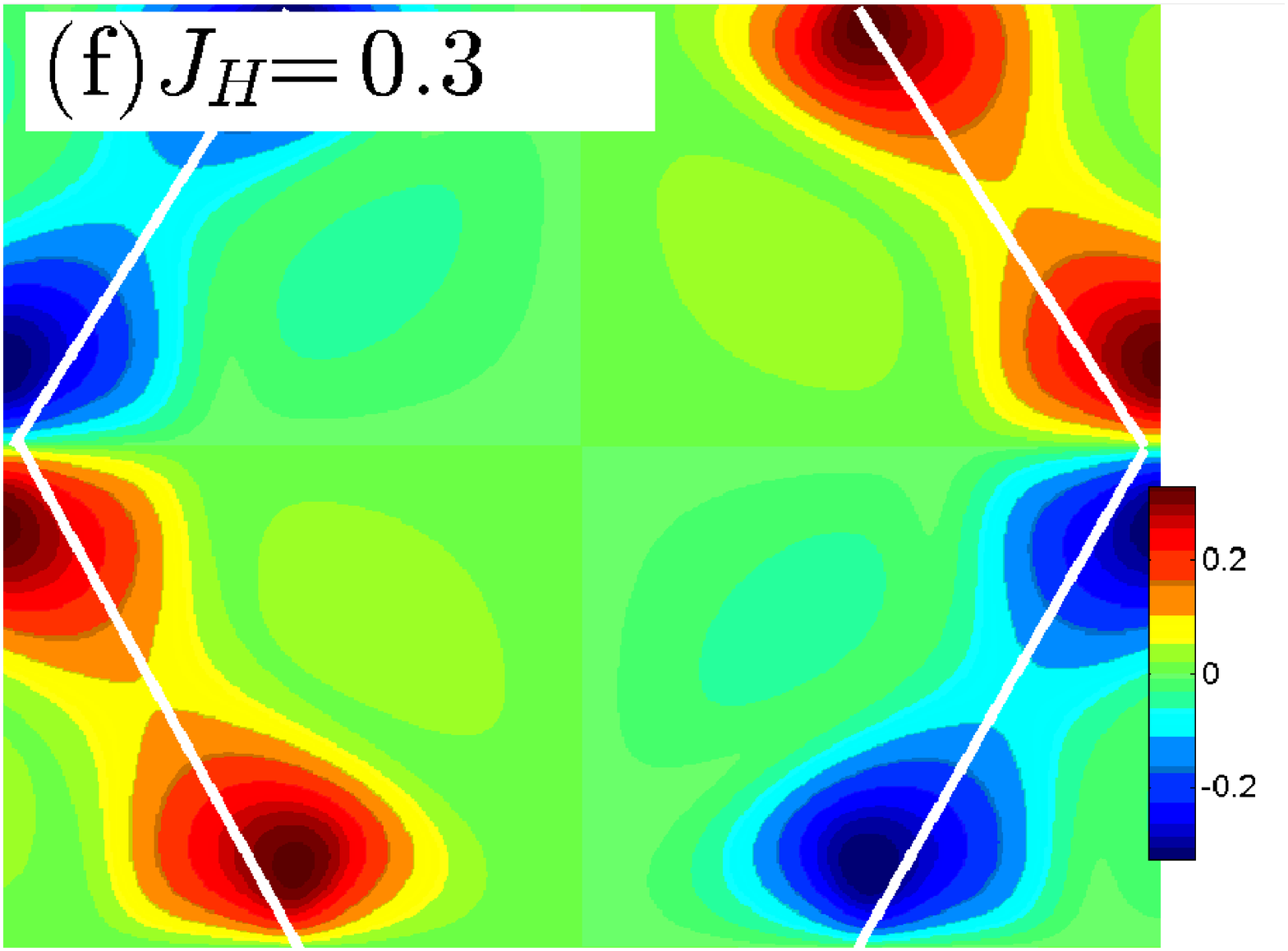}
\includegraphics[width = 0.32\columnwidth]{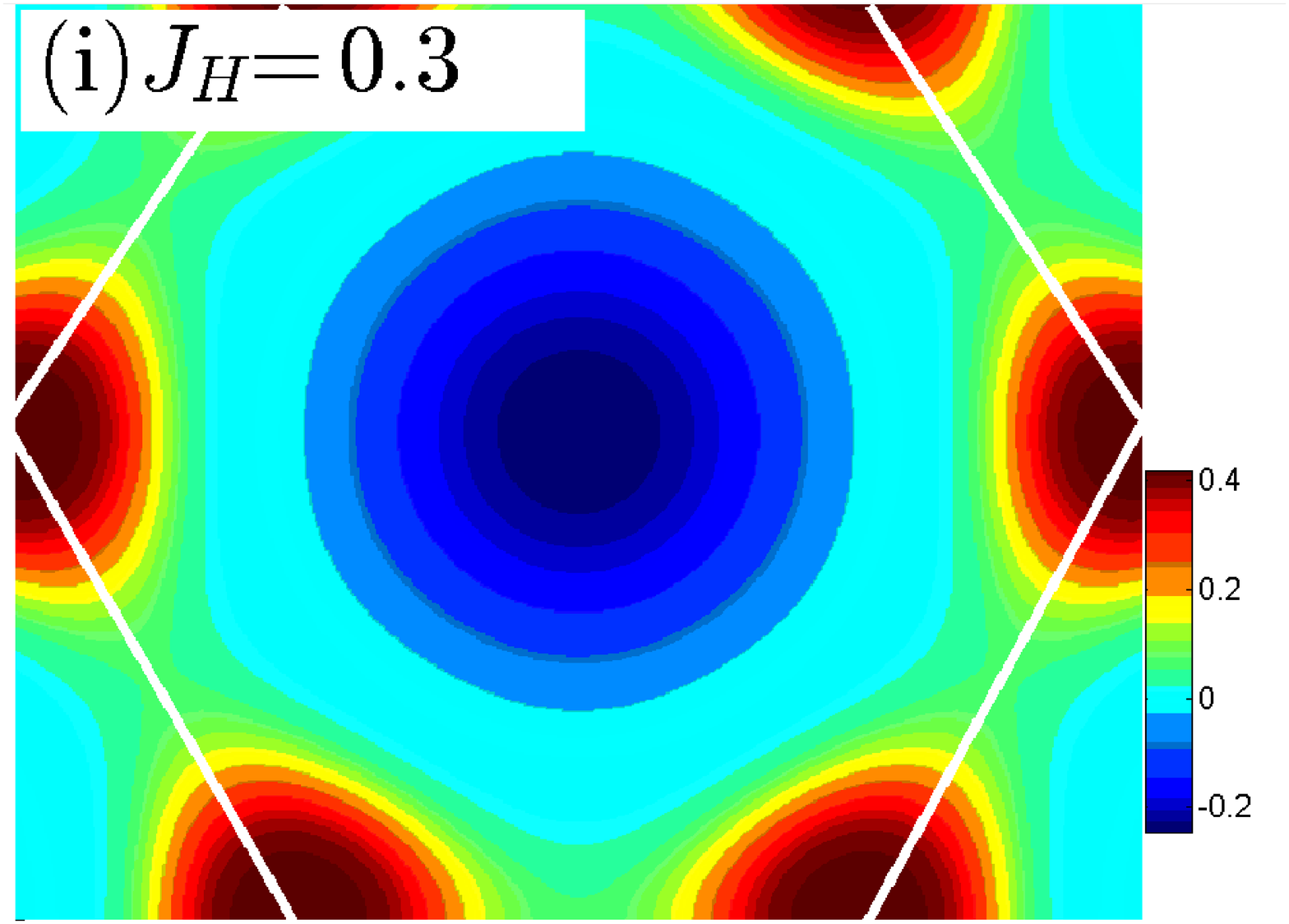}
\caption{\label{fig:pic6}The spinon pairing distribution for different $J_H$ in the first Brillouin zone marked by white lines. The first column [(a)-(c)] is the real parts of the chiral $d+id$-wave pairing functions and the second column [(d)-(f)] is the corresponding imaginary parts. The third column [(g)-(i)] is the extend $s$-wave pairing functions. $\Gamma$, K and M are high symmetric points of the triangular lattice. The parameters used are the same as those in Fig. \ref{fig:pic4}. }
\end{figure}

To further understand the superconducting instability, it is helpful to discuss the spinon pairing function in momentum space, which can be deduced by using the standard equation of motion method and the spectral theorem of the retarded Green's function, as also done in Ref. \cite{LIU2014}. The result is given by
\begin{equation}\label{eq:A15}
\left\langle f^{\dag}_{k\uparrow}f^{\dag}_{-k\downarrow}\right\rangle=
-\frac{J_H\Delta_{k}}{2\sqrt{2(E_{k1}+E_{k2})}}\left[1+\frac{\varepsilon^{2}_{k}}{E_{k2}}\right].
\end{equation}
The spinon pairing distribution for different $J_H$ is shown by the first two columns in Fig. \ref{fig:pic6} for the chiral $d+id$-wave pairing, in which the first column is the real part (with $d_{x^2-y^2}$ symmetry) and the second one is the imaginary part (with $d_{xy}$ symmetry), and the third column for the extended $s$-wave pairing. Comparison with these two cases, one notes that for small $J_H$, as shown in Fig. \ref{fig:pic6} (a),(d),(g), the pairing function of the chiral $d+id$-wave pairing is very weak but it is strong for the extended $s$-wave case. In this situation, the system prefers to be the extended $s$-wave pairing. With increasing $J_H$ up to 0.2292, these two cases have comparable pairing functions, and an energy gap begins to open in the case of the chiral $d+id$-wave pairing. This is the point at which the system prefers to be the chiral $d+id$-wave pairing. Further increasing $J_H$, the energy gap of the chiral $d+id$-wave pairing continues to increase, but it is still smaller than the extended $s$-wave pairing case. Therefore, for large $J_H$, the system has the chiral $d+id$-wave pairing.

\begin{figure}[tdp]
\centering
\includegraphics[width = 0.32\columnwidth]{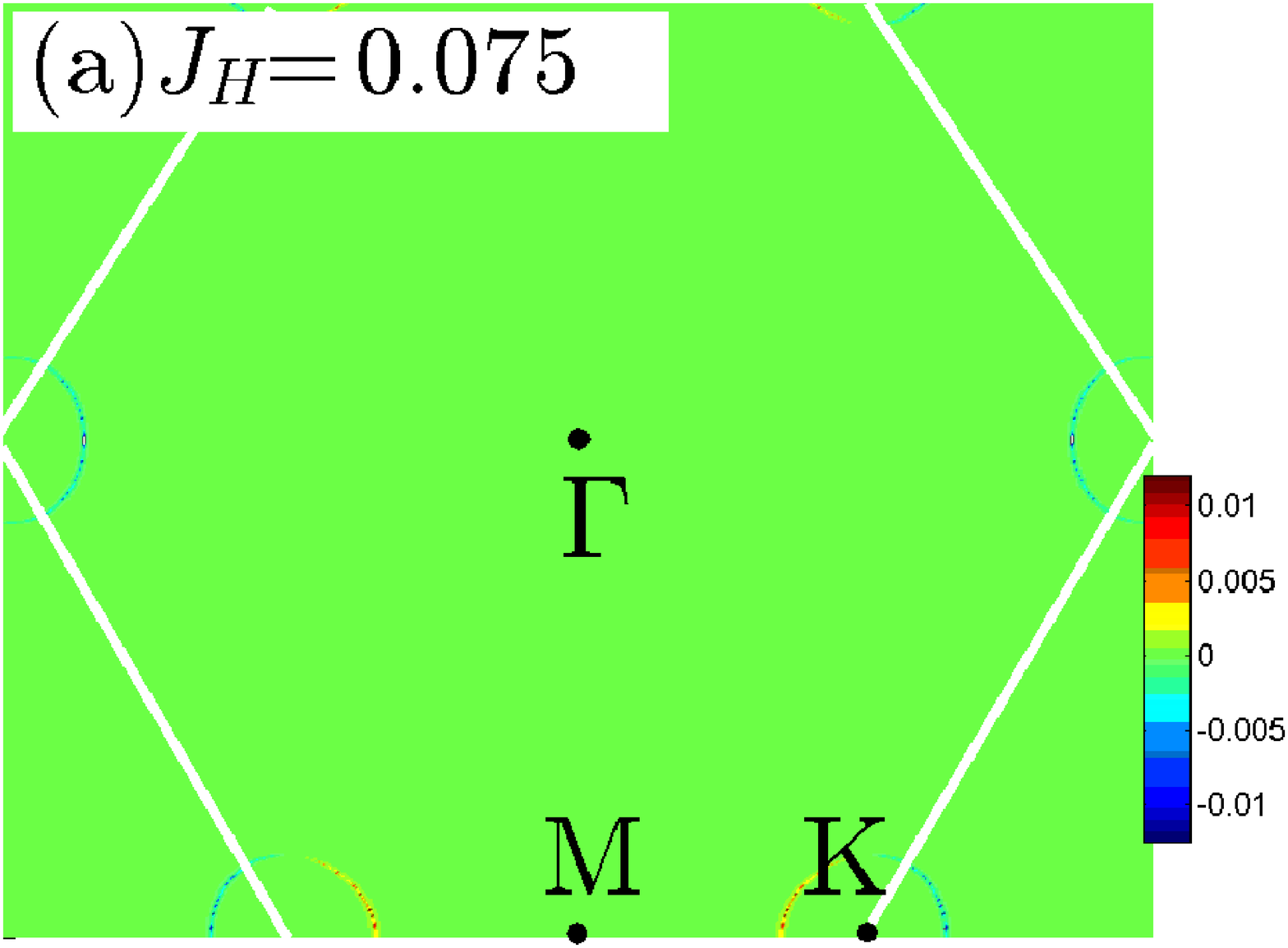}
\includegraphics[width = 0.32\columnwidth]{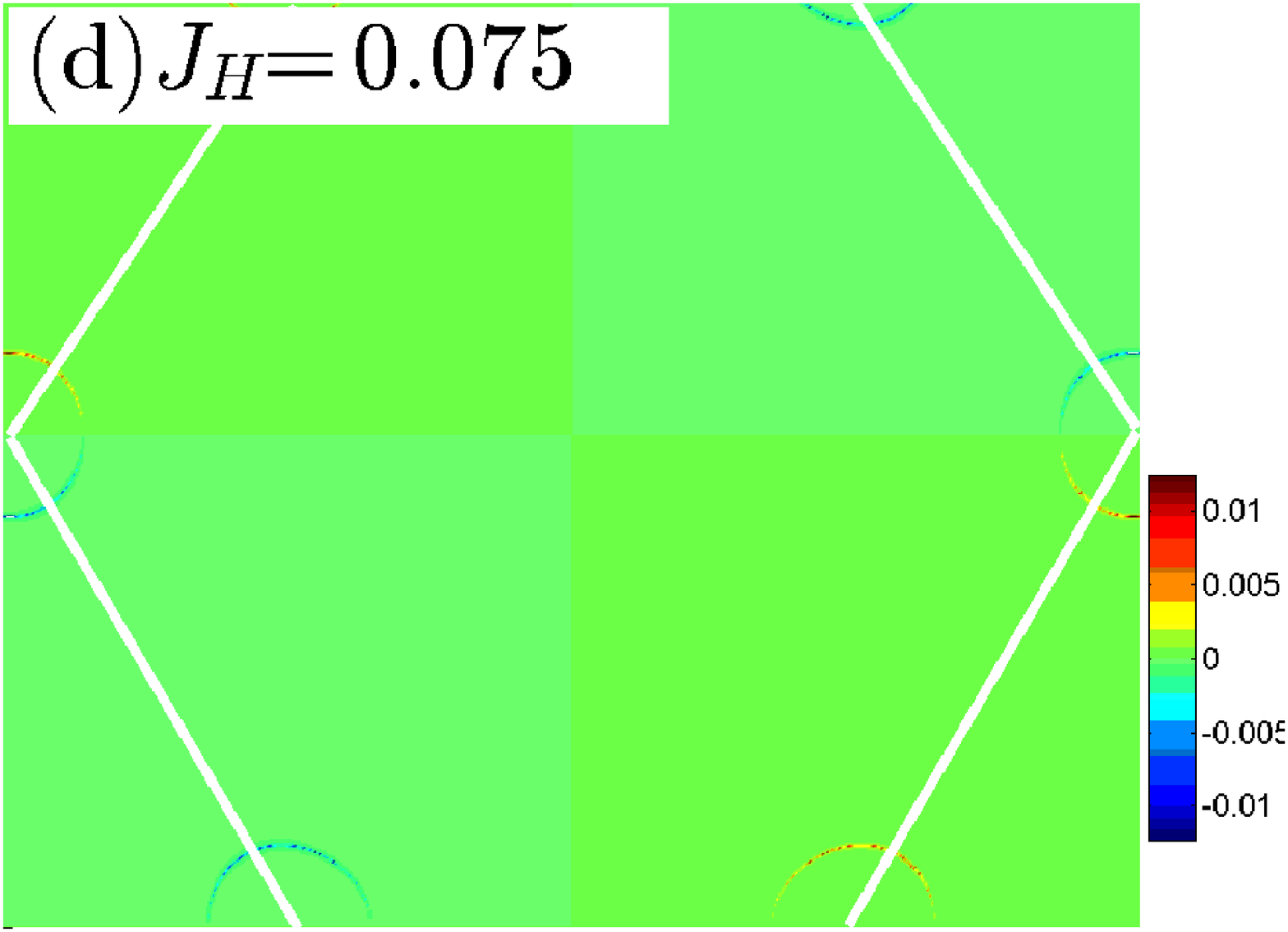}
\includegraphics[width = 0.32\columnwidth]{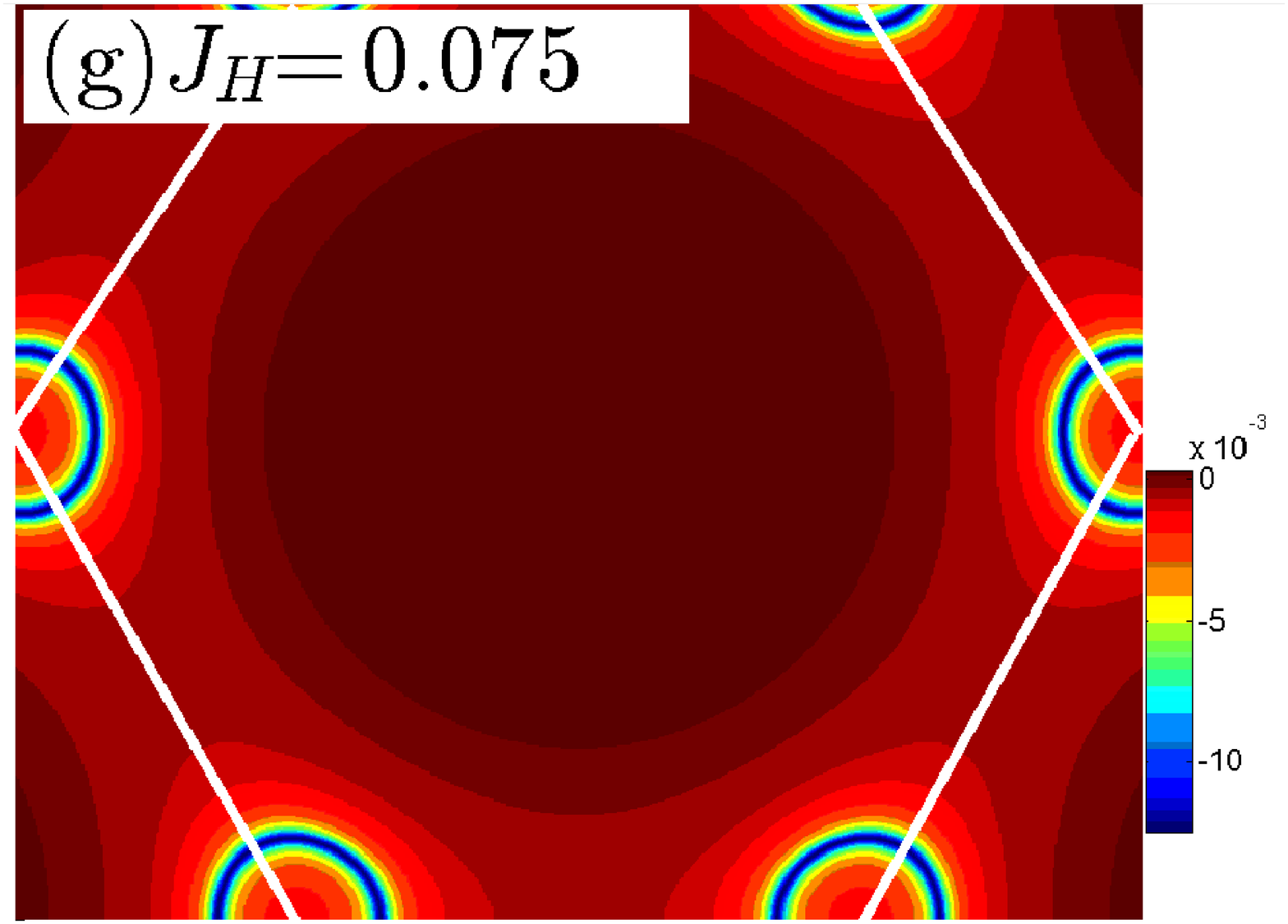}
\includegraphics[width = 0.32\columnwidth]{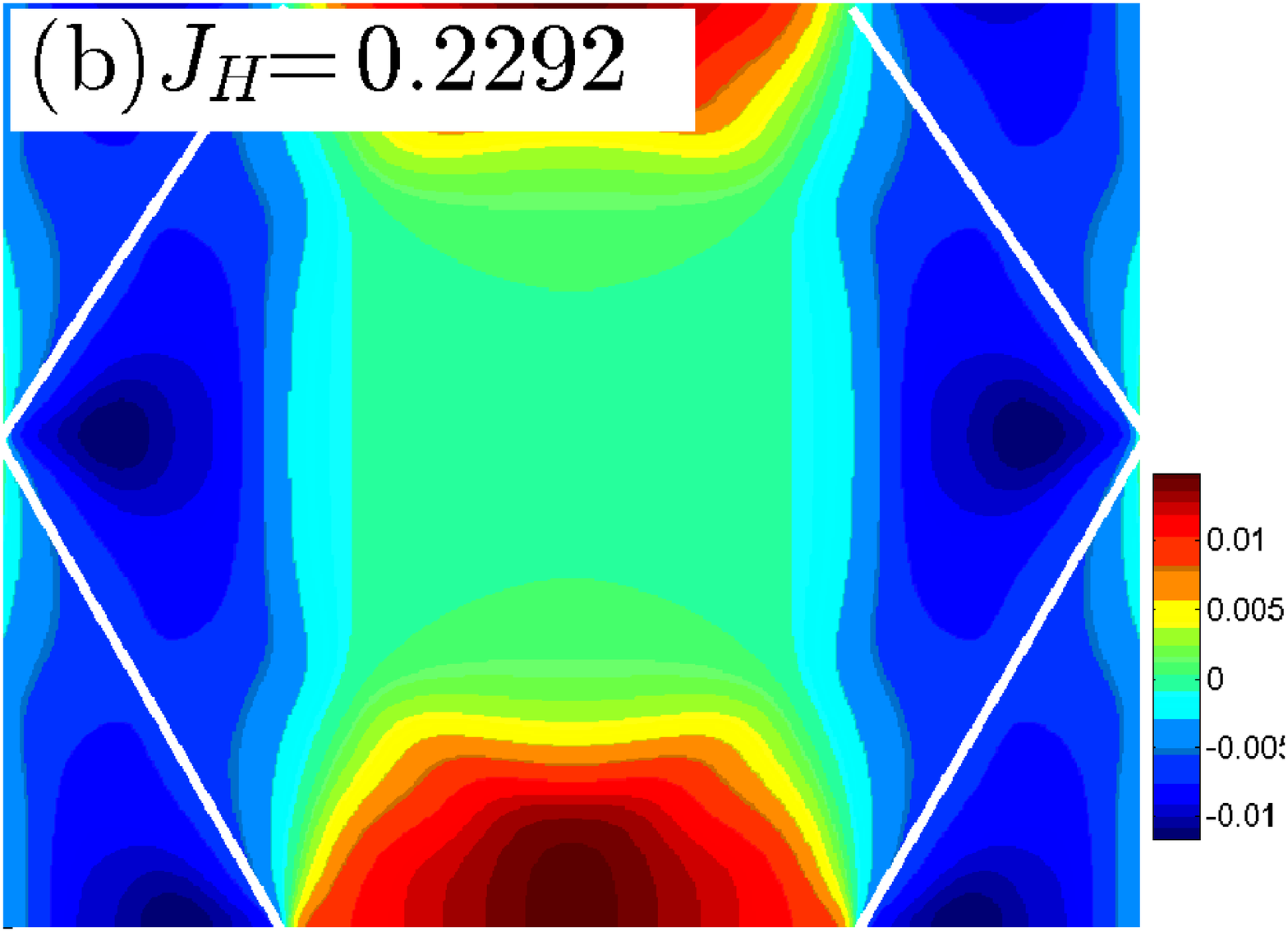}
\includegraphics[width = 0.32\columnwidth]{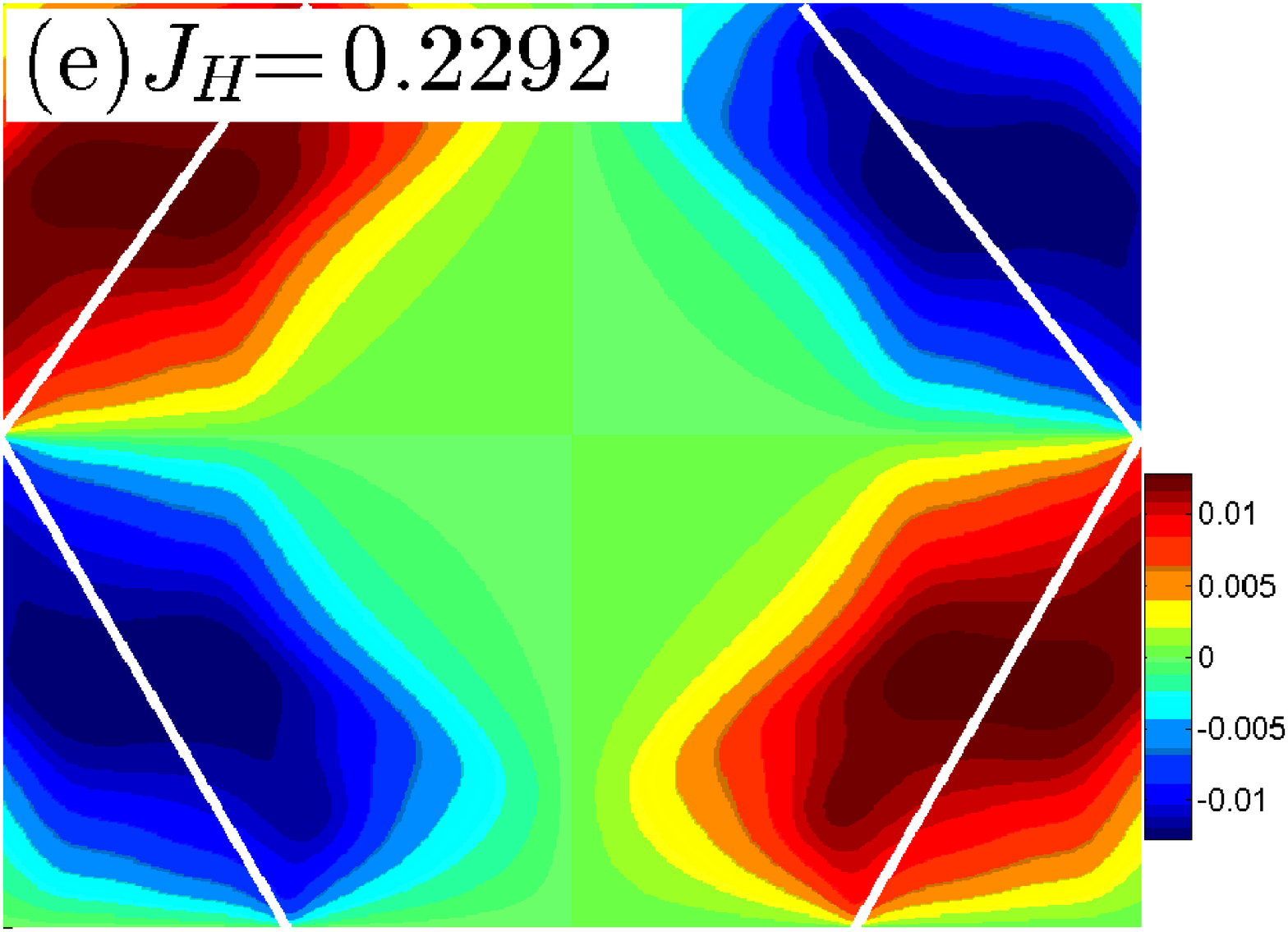}
\includegraphics[width = 0.32\columnwidth]{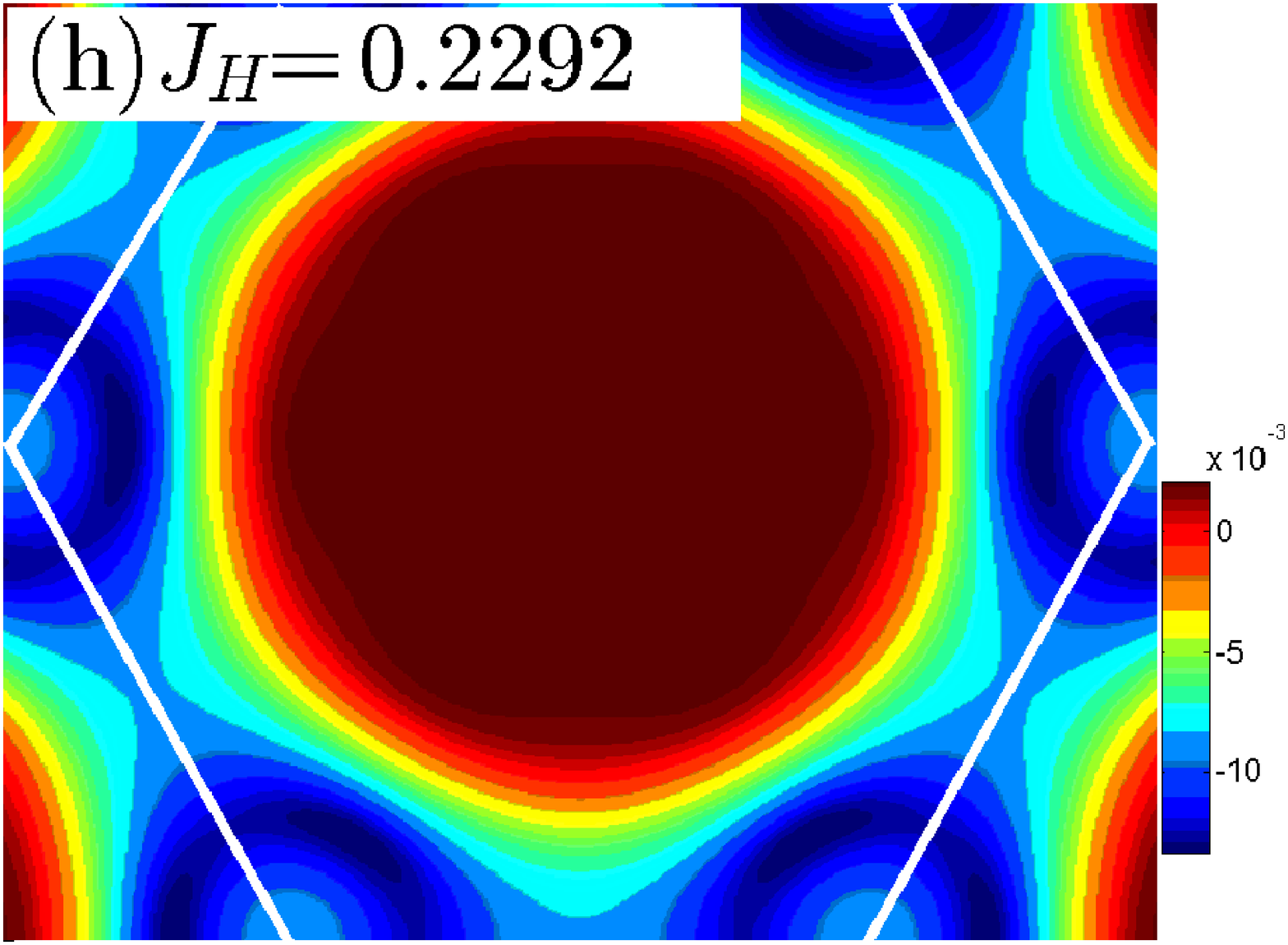}
\includegraphics[width = 0.32\columnwidth]{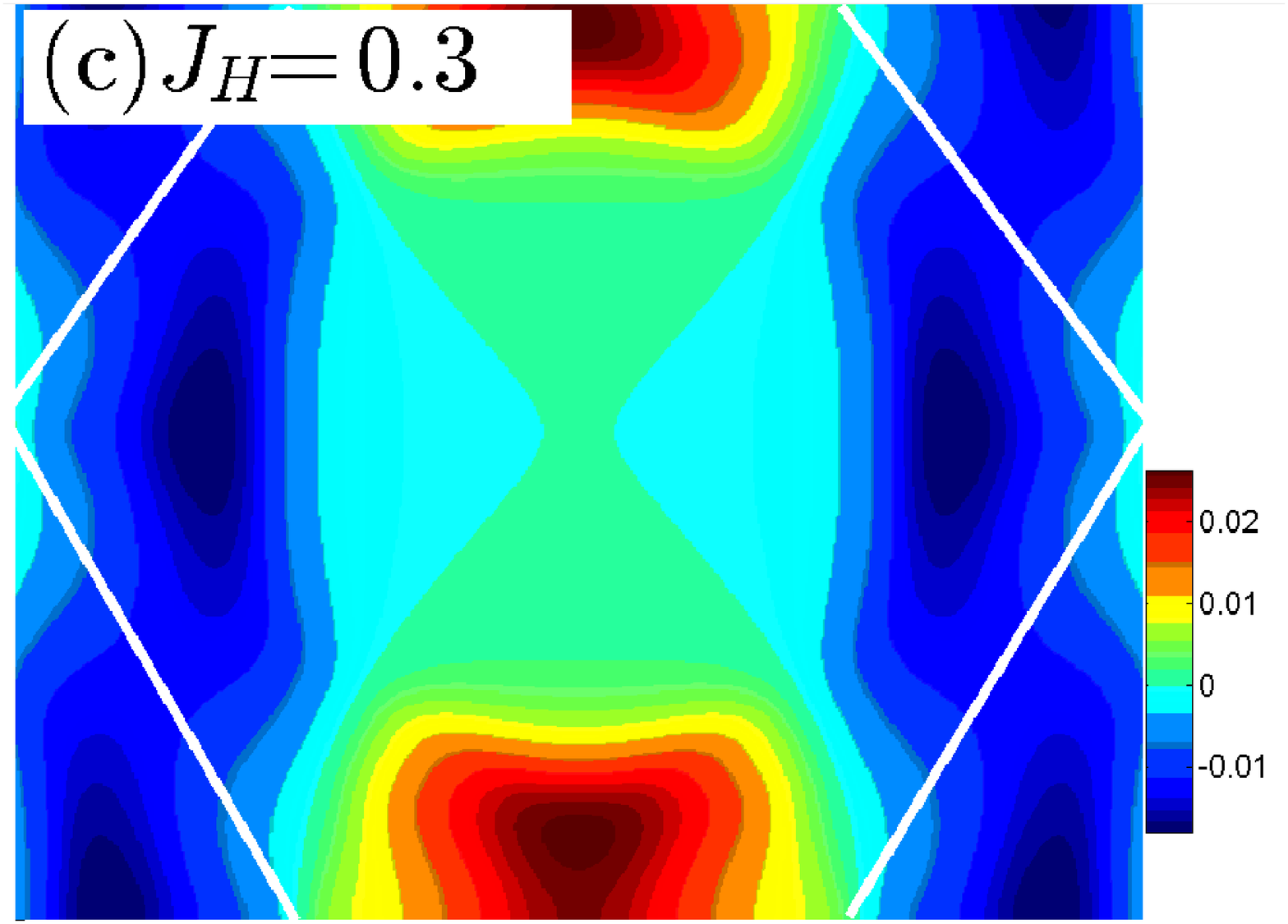}
\includegraphics[width = 0.32\columnwidth]{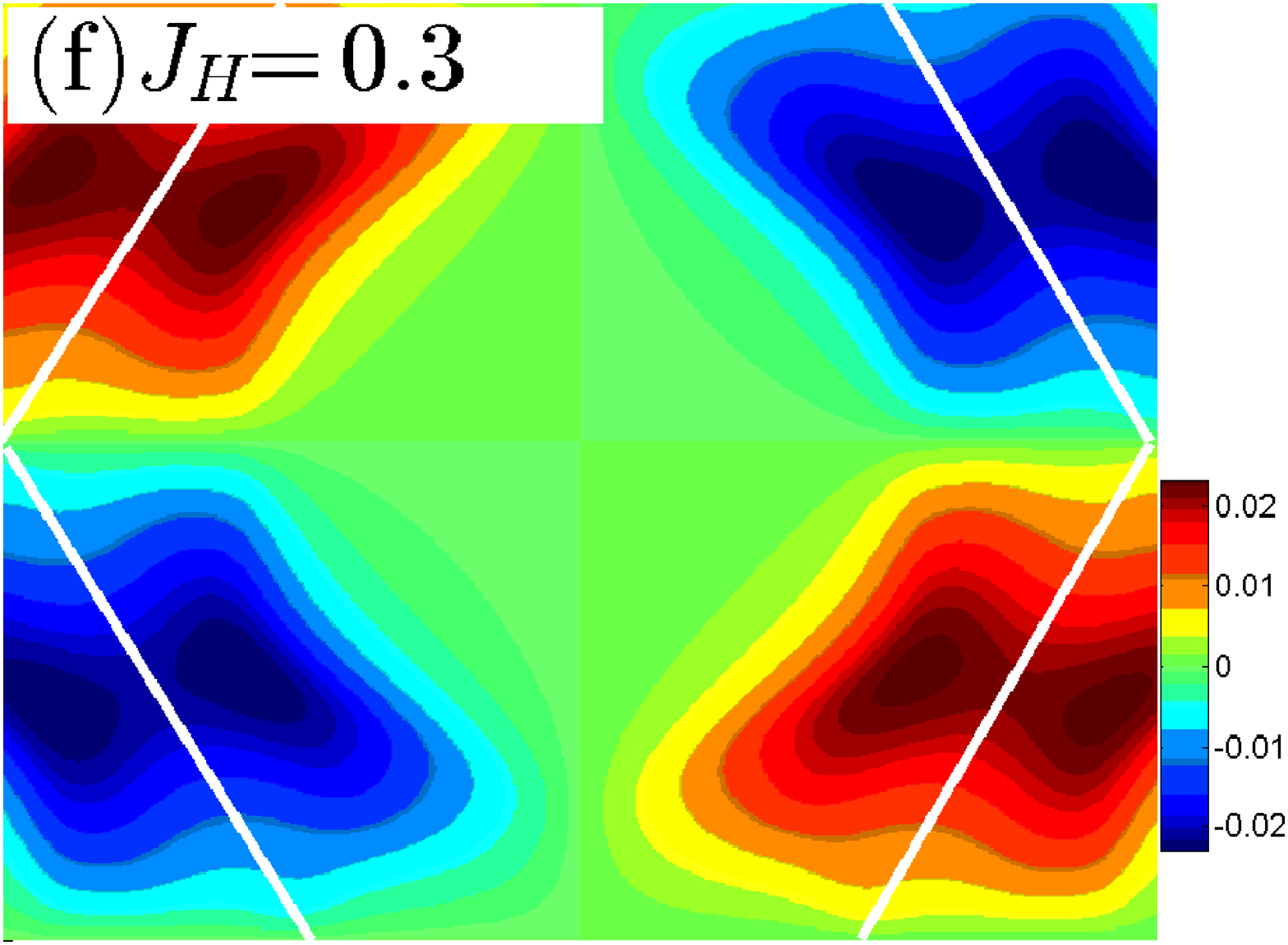}
\includegraphics[width = 0.32\columnwidth]{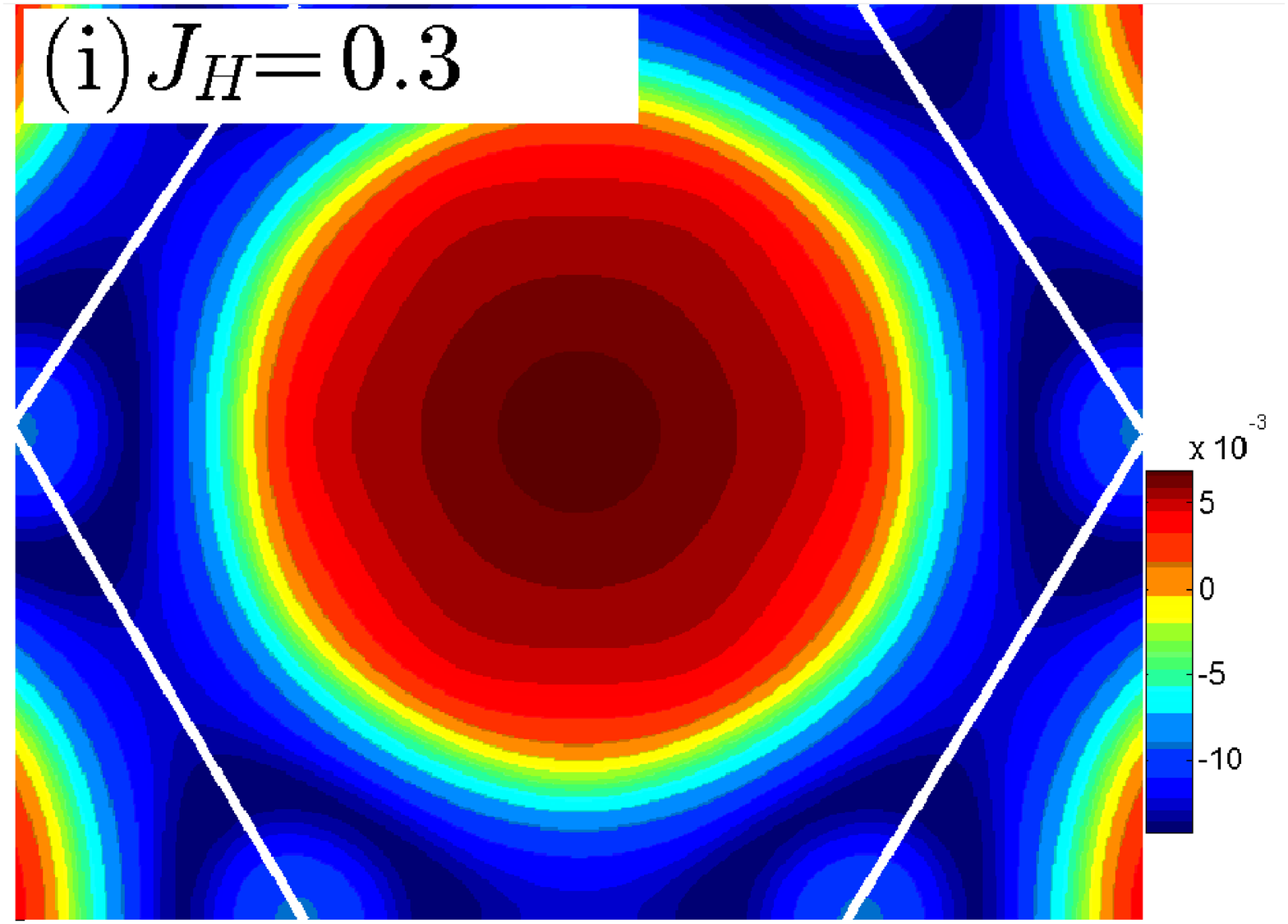}
\caption{\label{fig:pic8} The conduction electrons pairing distribution for different $J_H$ in the first Brillouin zone marked by white lines. The first column [(a)-(c)] is the real parts of the chiral $d+id$-wave pairing functions, the second column [(d)-(f)] is the corresponding imaginary parts. The third column [(g)-(i)] is the extend $s$-wave pairing functions. $\Gamma$, K and M are high symmetric points of the triangular lattice. The parameters used are the same as those in Fig. \ref{fig:pic4}. }
\end{figure}

We can also calculate the pairing functions for the conduction electrons induced by the Kondo coupling. It is given by
\begin{equation}
\left\langle c^{\dag}_{k\uparrow}c^{\dag}_{-k\downarrow}\right\rangle
=\frac{J_{H}J^{2}_{K}V^{2}\Delta_{k}}{8E_{k2}\sqrt{2(E_{k1}+E_{k2})}}.
\end{equation}
As shown in Fig.\ref{fig:pic8}, crudely speaking, the pairing function of conduction electrons have the similar pairing symmetry as spinon, but show a $\pi$-phase shift in comparison to the case of spinon pairing shown in Fig.\ref{fig:pic6}, which is consistent with the result on square lattice.\cite{Liu2012} It is noticed that the pairing intensity of conduction electrons is much smaller than its spinon counterpart, which reflects the fact that the pairing of conduction electron is induced by the preformed pairing of local electrons via non-vanishing Kondo screening.\cite{Coleman1989,Zhong2014}

\begin{figure}[h]
\centering
\includegraphics[width = 0.485\columnwidth]{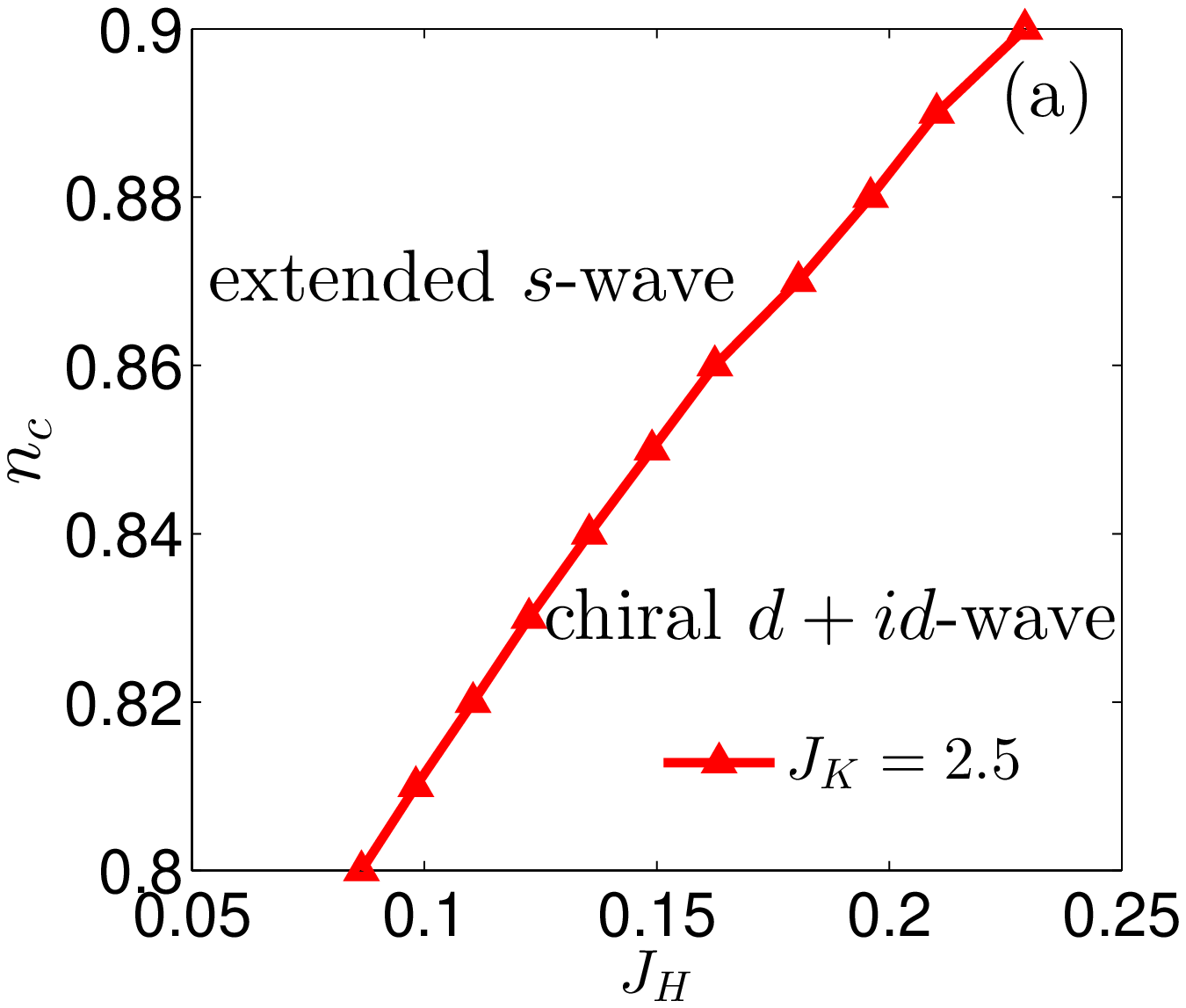}
\includegraphics[width = 0.485\columnwidth]{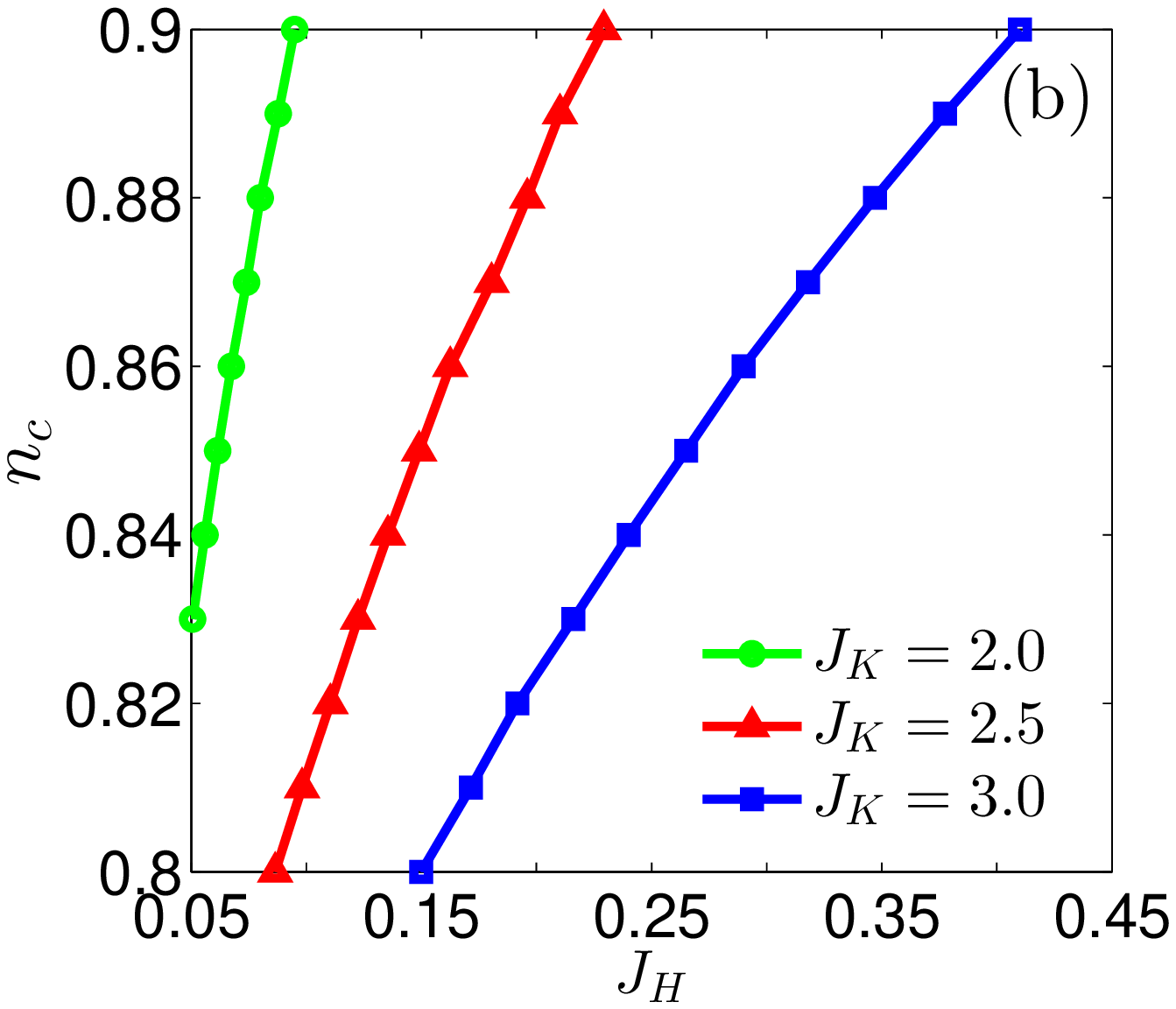}
\caption{\label{fig:pic9} (a) The phase diagram of the pairing symmetry for the KHM model on the triangular lattice in the $n_c-J_H$ plane for $J_K = 2.5$. (b) The phase boundary for different $J_K$. The parameters used are the same as those in Fig. \ref{fig:pic4}. }
\end{figure}

In the above discussion, we have fixed the concentration of conduction electrons by $n_c = 0.9$. In order to study other concentrations of the conduction electrons, we have obtained a phase diagram of the pairing symmetry for the KHM model on the triangular lattice, as shown in Fig. \ref{fig:pic9}(a) for $J_K = 2.5$. The phase diagram of pairing symmetry is separated by the cross points of the ground-state energy in Fig. \ref{fig:pic4}(a). Due to the smooth cross of these two lines and different slopes, one can conclude that the phase transition is a first-order. Therefore, we do not expect radical changes of thermal and transport behaviors near this transition point. For different $J_K$, the result is shown in Fig. \ref{fig:pic4}(b).

\begin{figure}[tdp]
\centering
\includegraphics[width = 0.485\columnwidth]{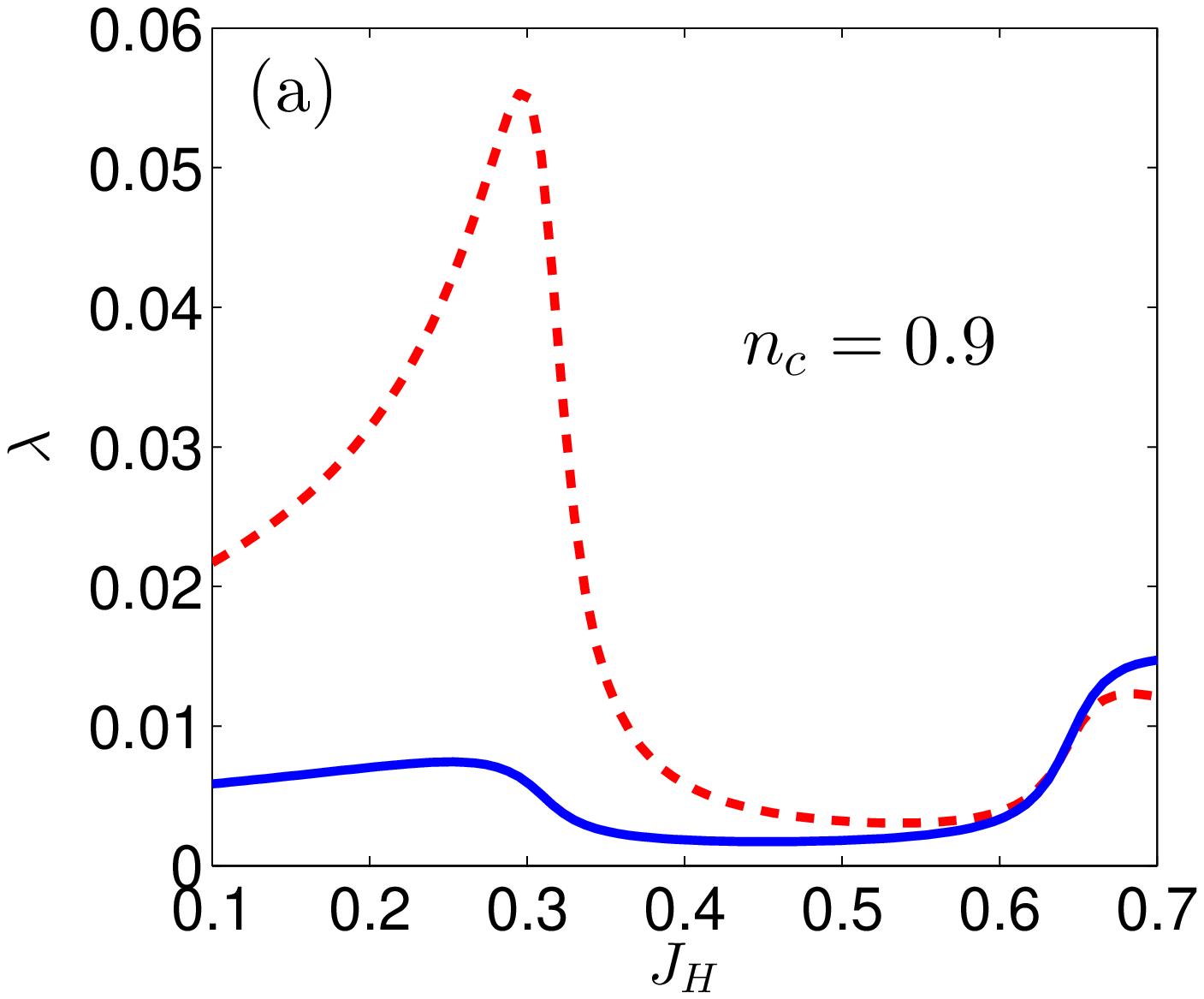}
\includegraphics[width = 0.485\columnwidth]{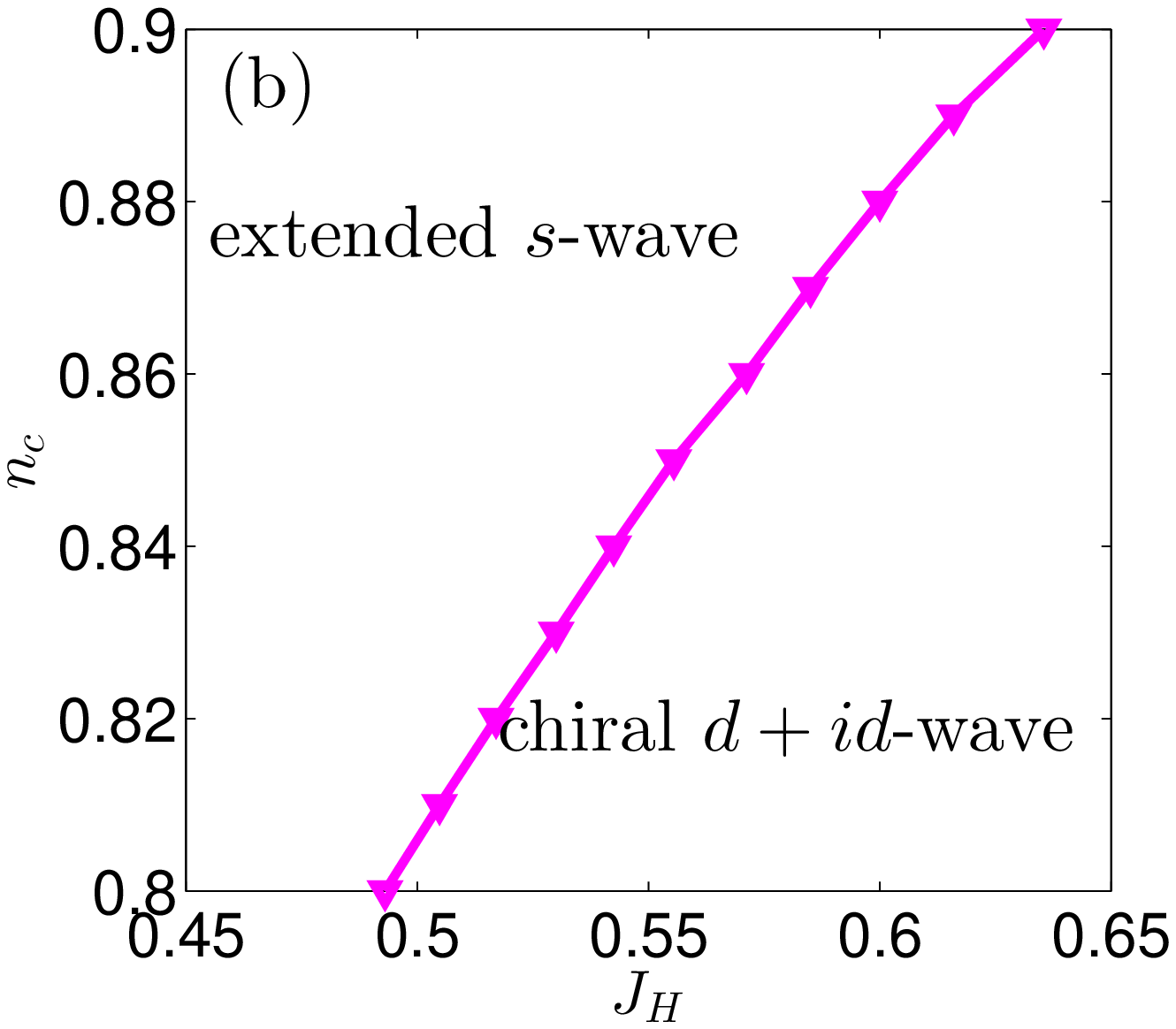}
\caption{\label{fig:pic11} (a) The integrated pairing strengths of the extended $s$-wave (red dashed line) and the chiral $d+id$-wave (blue solid line) pairing symmetry and (b) the phase diagram of the pairing symmetry in the $n_{c} - J_H$ plane. The parameters used are the same as those in Fig. \ref{fig:pic4}. }
\end{figure}

\begin{figure}[tdp]
\centering
\includegraphics[width = 0.485\columnwidth]{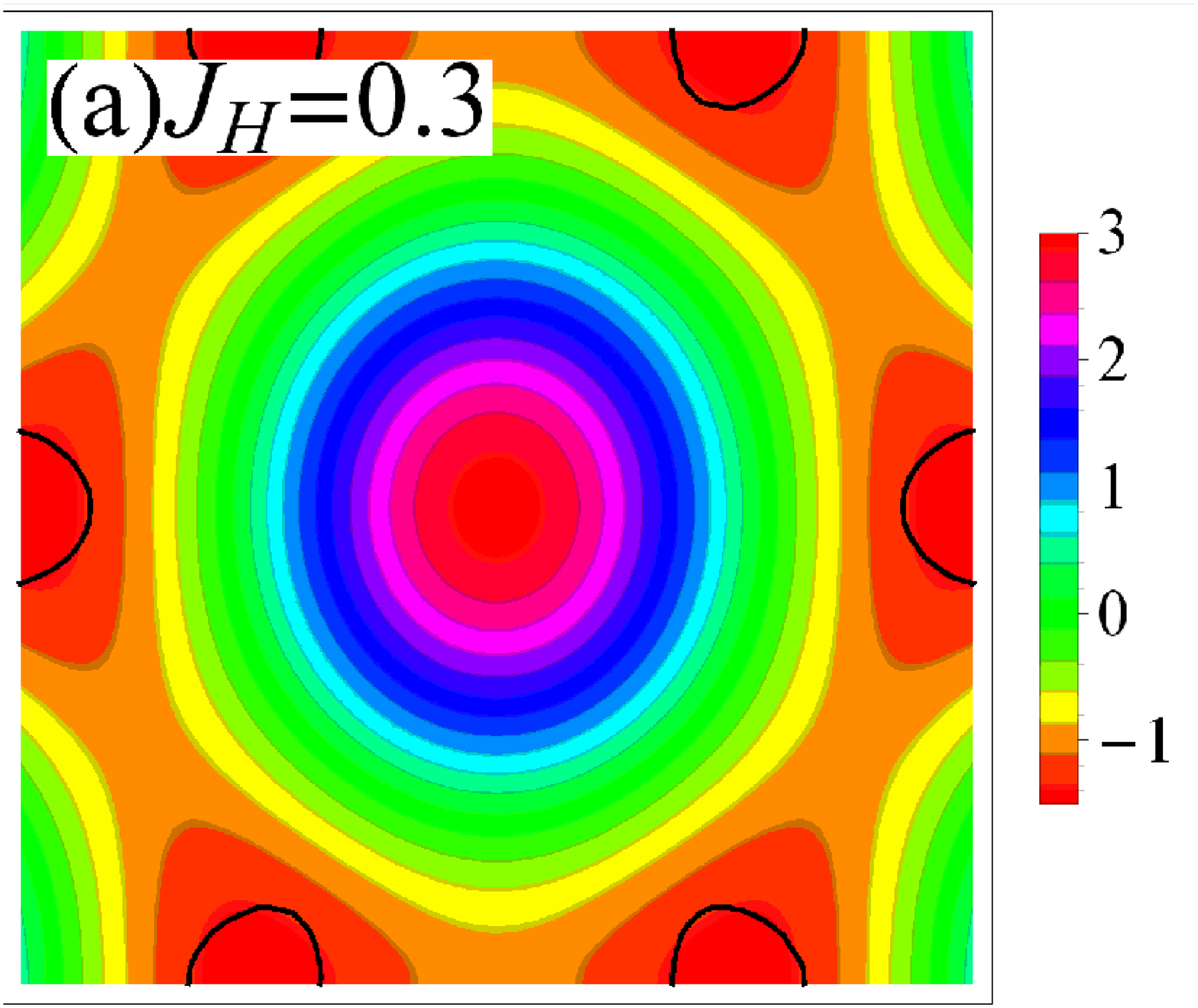}
\includegraphics[width = 0.485\columnwidth]{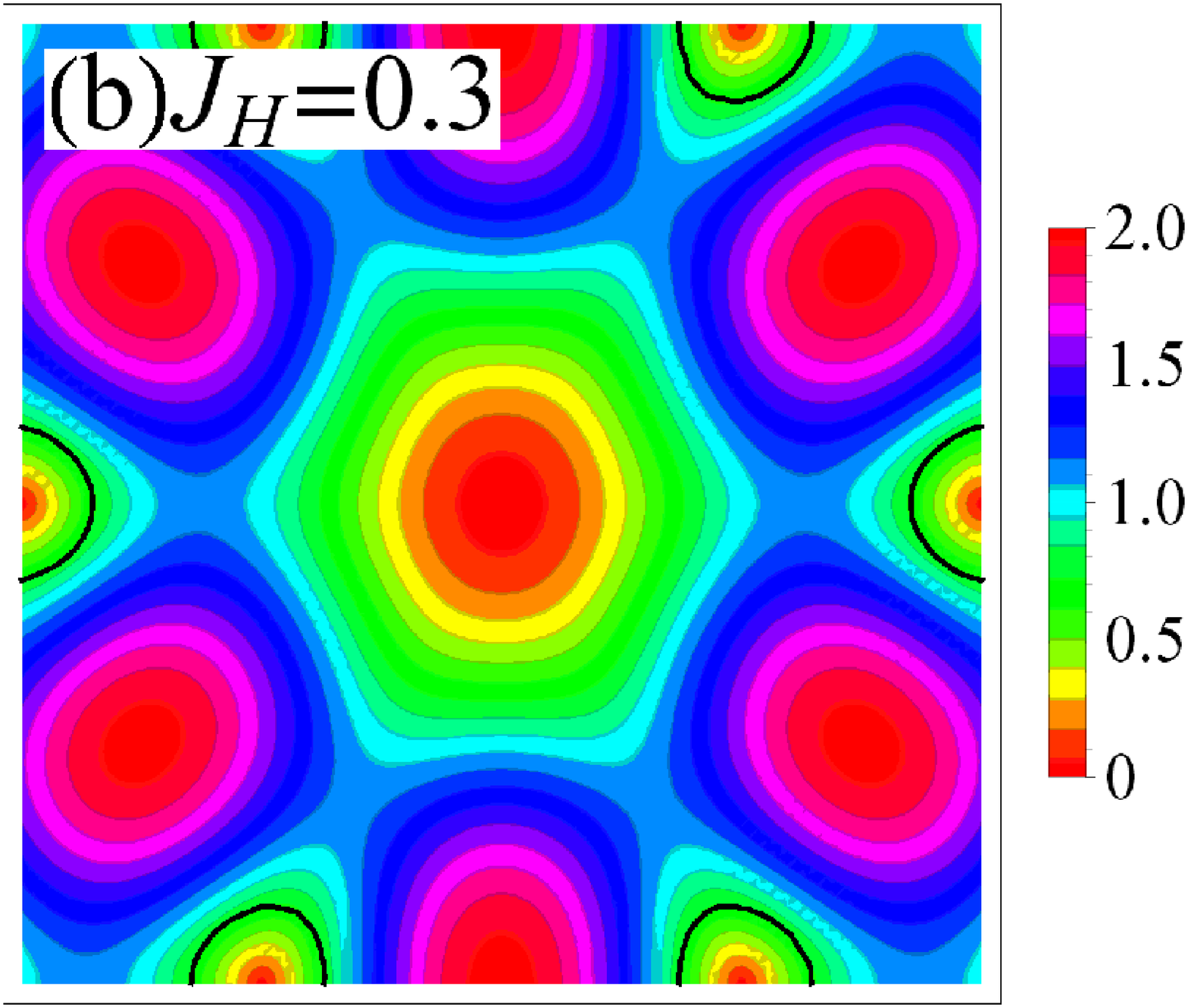}
\includegraphics[width = 0.485\columnwidth]{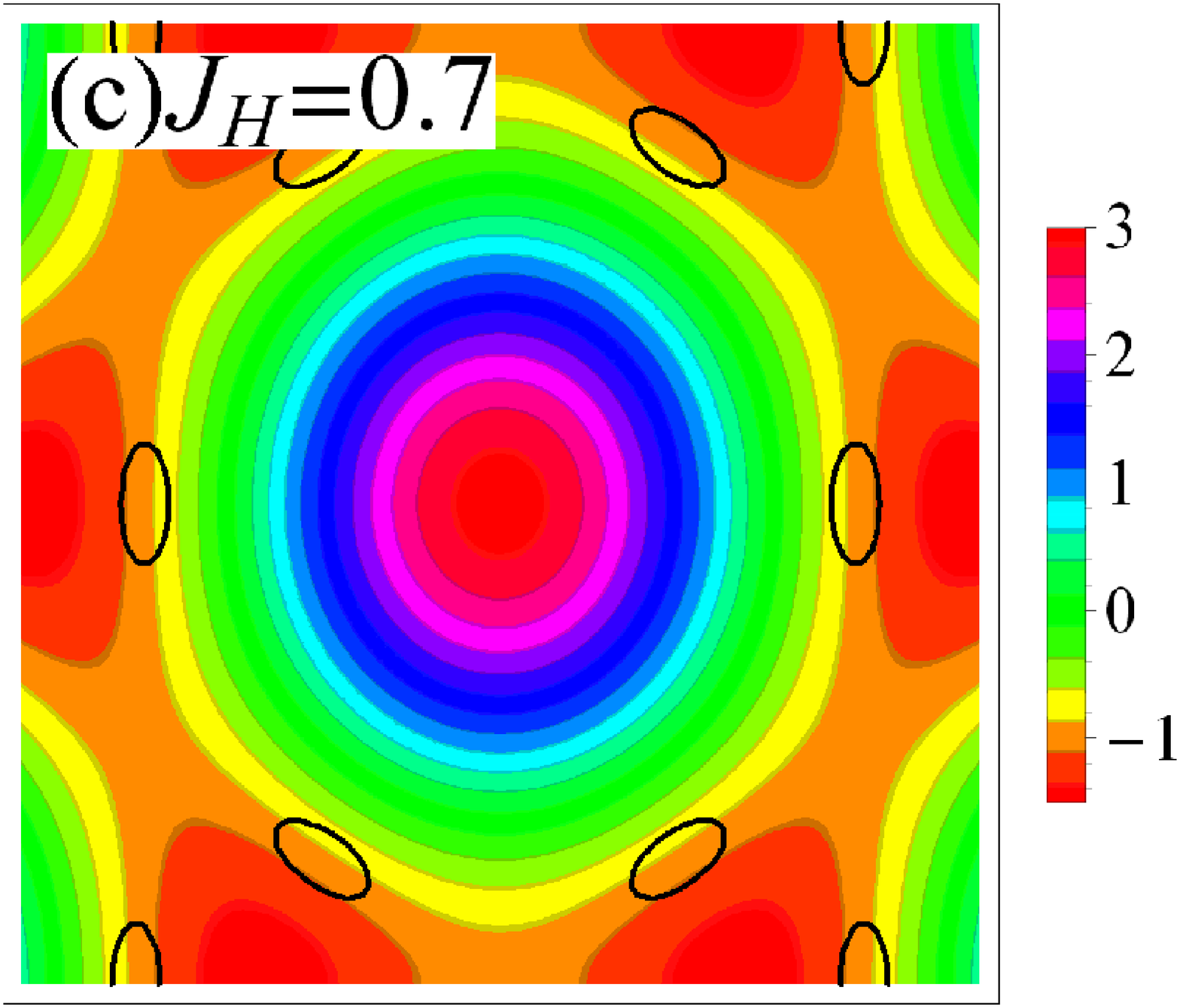}
\includegraphics[width = 0.485\columnwidth]{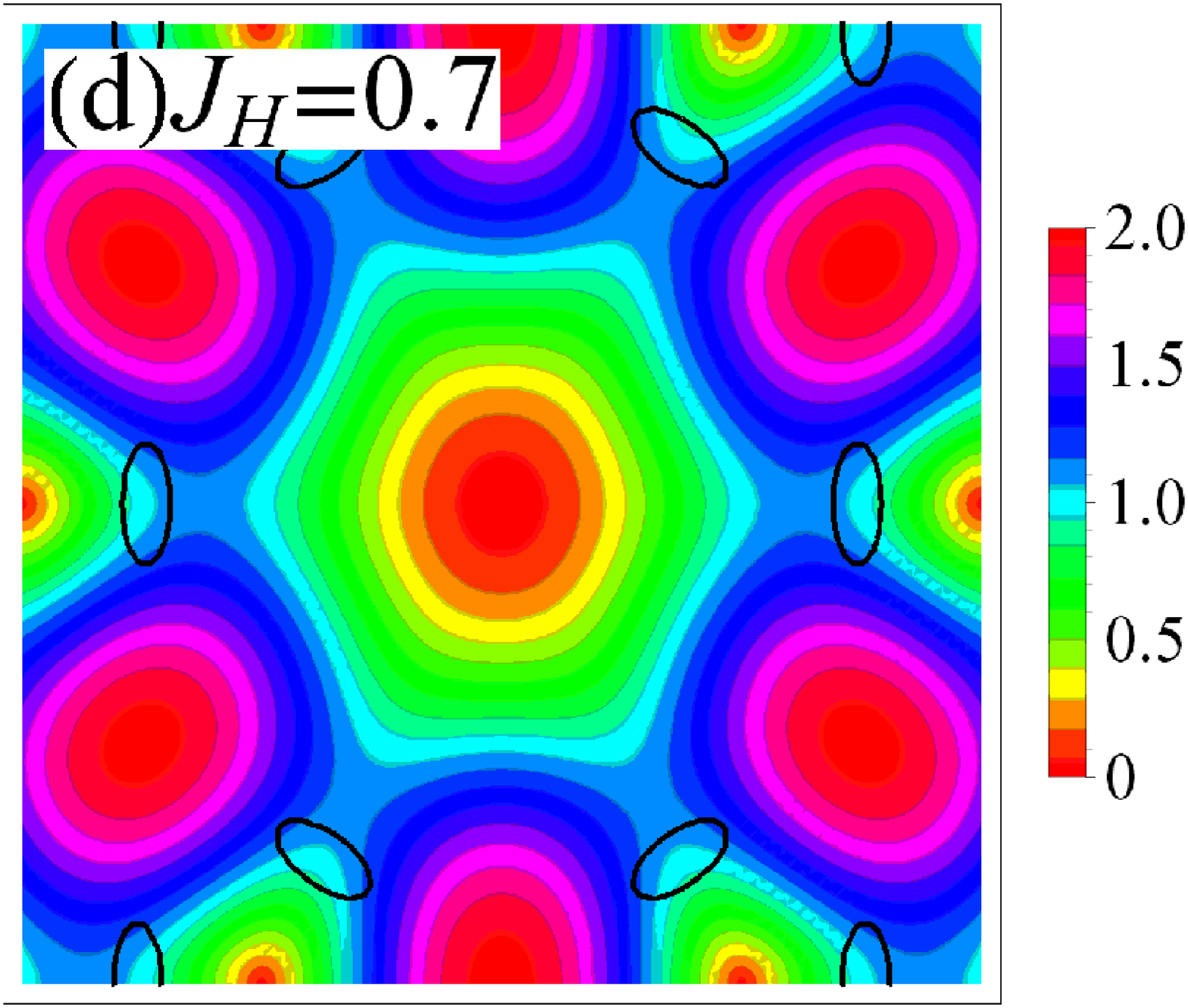}
\caption{\label{fig:pic10} The pairing strengths in the momentum space of extended $s$-wave pairing symmetry[(a) and (c)] and the chiral $d+id$-wave [(b) and (d)] for different Heisenberg antiferromagnetic interactions. The black curve shows the Fermi surface. Here $n_c = 0.9$ and the other parameters used are the same as those in Fig. \ref{fig:pic4}. }
\end{figure}

To understand the physics of the pairing symmetry transition, it is also useful to check the pairing strength, a quantity proposed by Hu and Ding in order to explain the resulting pairing symmetry on specific lattices.\cite{Hu2012} In the present case, the superconducting pairing strength is given by
\begin{equation}
\lambda=\sum_{k}|f_{k}|^{2}\left[\delta\left(E^{+}_{k}\right)+\delta\left(E^{-}_{k}\right)\right],
\end{equation}
where the form factor of the extended $s$-wave is $f^{s}_{k}=\cos(k_{x})-\cos\left(\frac{k_{x}}{2}\right)\cos\left(\frac{\sqrt{3}k_{y}}{2}\right)$, and that of the chiral $d+id$-wave is $f^{d+id}_{k}=\cos(k_{x})-i\sqrt{3}\sin\left(\frac{k_{x}}{2}\right)\sin\left(\frac{\sqrt{3}k_{y}}{2}\right)
-\cos\left(\frac{k_{x}}{2}\right)\cos\left(\frac{\sqrt{3}k_{y}}{2}\right)$. The normal state quasi-particle excitation spectra are defined as $E^{\pm}_{k}=\frac{1}{2}\left[(\varepsilon_{k}-\mu+\chi_{k})\right.$ $\left.\pm\sqrt{(\varepsilon_{k}-\mu-\chi_{k})^{2}+(J_{K}V)^{2}}\right]$.\cite{Hu2012} The result of the integrated pairing strength is shown in Fig. \ref{fig:pic11}(a). It is very clear that for small $J_H$ and $n_c = 0.9$, the pairing strength of the extended $s$-wave is quite larger than that of the chiral $d+id$-wave. Increasing $J_H$, the pairing strengths show a cross point around $J_H = 0.6355$, which is identified as a first-order transition. When one changes $n_c$, one obtains a phase diagram of the pairing symmetry according to the pairing strength, as shown in Fig. \ref{fig:pic11}(b), which shows qualitatively the similar trend as the one obtained by the ground-state energy calculations, though the Heisenberg antiferromagnetic interactions have quite different values since the methods are different, as explained below.

Fig. \ref{fig:pic10} shows the pairing strength in the momentum space and the corresponding Fermi surface at $n_c=0.9$ for different Heisenberg antiferromagnetic interactions. It is very clear that the pairing strength is quite sensitive to the shape of Fermi surface. On the contrary, the result from the ground-state energy is not so sensitive to the shape of the Fermi surface. Therefore, the phase diagram presented in Fig. \ref{fig:pic11} (b) can reflect more physics of the Fermi surface topology. In this sense, Fig. \ref{fig:pic11} (b) may be more practical for real systems, \cite{Kiesel2013} for example, the water-intercalated sodium cobaltates Na$_{x}$Co$_{2}\cdot y$H$_{2}$O.

\subsection{Effective single-band BCS description of superconducting states}
It is also interesting to note that although the KHM seems a two-band model, one may use an effective single-band BCS model to understand its basic features.\cite{Chan86JPCS-tsm,Miya86PRB-Sfm} The main point is that for most of heavy fermion superconductors, the superconducting state is believed to stem from a high temperature normal heavy Fermi liquid state, thus one can expect that the superconducting phase results from the pairing of renormalized heavy quasiparticles.

Explicitly, we can first diagonalize Eq.(\ref{eq:A10}) to obtain the heavy quasiparticles energy band using the Bogoliubov transformation $c_{k\sigma}=u_{k}\alpha_{k\sigma}-v_{k}\beta_{k\sigma}$, $f_{k\sigma}=v_{k}\alpha_{k\sigma}+u_{k}\beta_{k\sigma}$.
Here the $\alpha_{k\sigma}$ and $\beta_{k\sigma}$ are the quasiparticle operators, and the $u_{k}$ ($v_{k}$) is the coherence factor. The Hamiltonian in terms of the new basis becomes
\begin{equation}\label{eq:A16}
H=\sum_{k\sigma}\left(E_{k}^{+}\alpha^{\dag}_{k\sigma}\alpha_{k\sigma}+E_{k}^{-}\beta^{\dag}_{k\sigma}\beta_{k\sigma}\right).
\end{equation}
We have neglected the constant terms and set $\chi=0$ since we only consider the pairing order parameter in Eq. (\ref{eq:A17}) shown below. Then we add the pairing terms and project such terms into the renormalized quasiparticles basis to obtain the effective Hamiltonian as follows
\begin{eqnarray}\label{eq:A17}
&&H=\sum_{k\sigma}\left\{\left(E_{k}^{+}\alpha^{\dag}_{k\sigma}\alpha_{k\sigma}
+E_{k}^{-}\beta^{\dag}_{k\sigma}\beta_{k\sigma}\right)\right.\nonumber\\
&&+J_{H}\left[\Delta_{k}\left(v_{k}v_{-k}\alpha^{\dag}_{k\uparrow}\alpha^{\dag}_{-k\downarrow}+
u_{k}u_{-k}\beta^{\dag}_{k\uparrow}\beta^{\dag}_{-k\downarrow}\right)+h.c.\right]\nonumber\\
&&\left.+J_{H}\left[\Delta_{k}\left(v_{k}u_{-k}\alpha^{\dag}_{k\uparrow}\beta^{\dag}_{-k\downarrow}+
u_{k}v_{-k}\beta^{\dag}_{k\uparrow}\alpha^{\dag}_{-k\downarrow}\right)+h.c.\right]\right\}\nonumber.\\
\end{eqnarray}
Since the Fermi surface lies in the $\beta$-quasiparticle energy band, the pairing mainly appears in this band. Neglecting the $\alpha$-band pairing and the inter-band pairing, we can get an effective single-band BCS model for the heavy fermion superconductivity.
\begin{equation}\label{eq:A18}
H=\sum_{k\sigma} \left[ E_{k}^{-}\beta^{\dag}_{k\sigma}\beta_{k\sigma} + \left(\Delta_{k}u_{k}u_{-k}\beta^{\dag}_{k\uparrow}\beta^{\dag}_{-k\downarrow}+h.c.\right)\right].
\end{equation}
For this effective single-band model, the important transport measurement on the inverse squared magnetic penetration depth $\lambda$ will have $\frac{1}{\lambda^{2}}\propto\frac{n_{tot}}{m^{\star}}$ with $n_{tot}=n_{c}+1$ and $m^{\star}$ being the effective mass of the heavy quasiparticle and $n_{c}$ is the concentration of conduction electrons. The superfluid density experiment would observe the expected exponent behavior at low temperature compared to the superconducting transition temperature since both the extended $s$-wave and the chiral $d+id$-wave are gapped and have no nodes for the present triangular lattice. In addition, one may also expect the optical conductance is gapped due to the pairing gap. However, it should be emphasized that both magnetic penetration depth and optical conductance experiments are not able to distinguish the pairing symmetry between the extended $s$-wave and the chiral $d+id$-wave. Therefore, a phase sensitive experiment to confirm the true pairing symmetry is needed if some real-life heavy fermion materials are indeed described by the present models with triangular lattice.

\section{Conclusion and perspective}\label{sec:3}
In summary, we have investigated the KHM on triangular lattice with the fermionic large-N mean-field theory. A qualitatively same phase diagram of the pairing symmetry has been obtained by comparing the ground-state energy and analyzing the pairing strength. At small Heisenberg antiferromagnetic interaction and large concentration of conduction electrons, the system prefers to an extended $s$-wave pairing symmetry and on the contrary, the system favors the chiral $d+id$-wave pairing symmetry. The phase transition between these two pairing symmetries is found to be first-order.

In very recent years, the issue of the pairing symmetry of the heavy fermion systems becomes quite active due to some experimental and theoretical advances. For example, the specific-heat measurements in CeCu$_2$Si$_2$ along with its linear dependence as a function of magnetic field and the absence of oscillations in the field angle suggest that this material may have an $s$-wave pairing symmetry. \cite{Kittaka2014} The $s$-wave pairing symmetry has also been predicted by a first-principle calculation in CeCu$_2$Si$_2$. \cite{Ikeda2015} For another heavy fermion compound CeCoIn$_5$, the London penetration depth measurements also suggest that a $Yb$ substitution can change the pairing symmetry from nodal to nodeless, which is closely related to the Fermi surface topology change. \cite{Kim2015} When it is still debated what pairing symmetry the heavy fermion superconductivity is, \cite{Kim2015,Erten2015,Masuda2015} our work provides a possible scenario to understand the pairing symmetry in heavy fermion superconductors. To test our result, it is hoped that future experiments can discover some heavy fermion compounds with the geometrical frustrated triangular lattice structure, which is highly desirable from the interest in unconventional superconductivity and the general frustration/interaction heavy fermion phase diagram.\cite{Coleman2010}

\section*{Acknowledgements}
The work is supported partly by NSFC, PCSIRT (Grant No. IRT1251), the Fundamental Research Funds for the Central Universities and the national program for basic research of China.

\end{document}